\documentclass[12pt]{article}
\usepackage{graphicx}
\usepackage{amsfonts}
\usepackage{amsmath}
\usepackage{mathrsfs}
\def\beq{\begin{equation}}
\def\eeq{\end{equation}}
\def\bea{\begin{eqnarray}}
\def\eea{\end{eqnarray}}
\def\nn{\nonumber}
\def\ba{\begin{array}}
\def\ea{\end{array}}

\def\v{\vert}
\def\l{\langle}
\def\r{\rangle}
\def\one{1\hskip -1mm{\rm l}}

\begin{document}

\begin{center}
{\large \bf \sf
Exact partition function of $SU(m|n)$ supersymmetric \\
Haldane-Shastry spin chain }

\vspace{1.3cm}

{\sf B. Basu-Mallick\footnote{ 
e-mail address: bireswar.basumallick@saha.ac.in}
and Nilanjan Bondyopadhaya\footnote{e-mail address:
nilanjan.bondyopadhaya@saha.ac.in } }

\bigskip

{\em Theory Group, \\
Saha Institute of Nuclear Physics, \\
1/AF Bidhan Nagar, Kolkata 700 064, India } \\
\bigskip

\end{center}

\vspace {2 cm}
\baselineskip=18pt
\noindent {\bf Abstract }

By taking the freezing limit of a spin Calogero-Sutherland model
containing `anyon like' representation of the permutation algebra, 
we derive the exact partition function of $SU(m|n)$ supersymmetric
Haldane-Shastry (HS) spin chain. 
This partition function allows us to study global properties 
of the spectrum like level density distribution and nearest neighbour
spacing distribution. It is found that, for supersymmetric HS spin chains 
with large number of lattice sites,  continuous part of the 
energy level density obeys Gaussian distribution with a 
high degree of accuracy. The mean value and standard deviation
of such Gaussian distribution can be calculated exactly.
We also conjecture that the partition function of supersymmetric HS spin 
chain satisfies a duality relation under the exchange of 
bosonic and fermionic spin degrees of freedom. 

\baselineskip=16pt
\vspace {.6 cm}

\vspace {.1 cm}

\newpage 
\baselineskip=18pt
\noindent \section {Introduction }
\renewcommand{\theequation}{1.{\arabic{equation}}}
\setcounter{equation}{0}

\medskip

Haldane-Shastry (HS) spin chain is 
a well known quantum integrable model, where equally spaced 
spins on a circle interact with each other through  
pairwise exchange interactions inversely proportional
to the square of their chord distances. 
Study of such HS spin-$\frac{1}{2}$ chain with long-range interaction
was originally motivated from the fact that the exact
ground state wavefunction of this model coincides with the 
$U \rightarrow \infty$ limit of Gutzwiller's
variational wave function for the Hubbard model,
and also with the one-dimensional version 
of the resonating valence bond state proposed by Anderson [1,2].
Remarkably,  HS spin chain can be explicitly solved
in much greater detail than integrable 
 spin chains with short-range interactions, has a Yangian 
quantum group symmetry and
interestingly shares many of the characteristics of
an ideal gas, but with fractional statistics [3-5].  
The Hamiltonian of $SU(m)$ HS model with $N$ number of lattice sites 
is given by
\beq
 H=\frac{1}{2}\sum_{1\leq j <k \leq N}
 \frac{(1+P_{jk})}{\sin^2(\xi_j - \xi_k )}  \, ,
\label {a1}
\eeq
where $\xi_j=j\pi / N$ and $P_{jk}$ is the exchange operator
interchanging the `spins' (taking $m$ possible values) on $j$-th and
$k$-th lattice sites.

By using the motif representations associated with
$Y(gl_m)$ Yangian symmetry of HS spin chain (\ref{a1}), 
one can find out its complete spectrum including the degeneracy factor 
for each energy level [6-8]. 
However, in practice, the computation of such degeneracy factors
becomes very cumbersome for $m>2$ and large values of $N$. 
Therefore, it is difficult to express 
 the partition function of HS spin chain in a simple form 
(for arbitrary values of $N$ and $m$) with the help of motif representations. 
Due to this reason, it is worthwhile to explore other approaches 
for calculating the partition function of HS spin chain. 
In fact, a rather simple expression for the exact partition function 
of $SU(m)$ HS spin chain has been obtained recently [9]
by applying the so called freezing trick [10-12].
This  freezing trick utilizes the connection between 
$SU(m)$ HS spin Hamiltonian 
and $SU(m)$ spin Calogero-Sutherland (CS) model which has dynamical 
 as well as spin degrees of freedom. More precisely, 
one takes the strong coupling limit of spin CS Hamiltonian, 
so that the particles freeze at their classical equilibrium positions 
of the scalar part of the potential and
spins get decoupled from the dynamical degree of freedom.
As a result, one can derive the partition function of HS spin chain
by `modding out' the partition function of spinless CS model from that of
the spin CS model.  By using this partition function
of $SU(m)$ HS spin chain, it is possible to study the energy level density
distribution and the nearest neighbour spacing distribution 
for fairly large values of $N$ [9]. Interestingly, 
it has been found that,  the continuous part of such 
energy level density follows Gaussian distribution to
a high degree of approximation. 

In this context it may be noted that,  there
exists a  $SU(m|n)$ supersymmetric extension of HS spin chain [3], where 
each site is occupied by either one of the $m$
type of bosonic states or one of the 
$n$ type of fermionic states.  Such supersymmetric spin chains
play an important role in describing some
correlated systems of condensed matter physics, 
where holes moving in the dynamical background of spin behave as 
bosons and spin-$\frac{1}{2}$ electrons behave as fermions [13]. It is worth
noting that the supersymmetric $SU(m|n)$ HS spin chain exhibits 
$Y(gl_{(m|n)})$ super-Yangian symmetry [3],  
which is also the quantum group symmetry of supersymmetric
$SU(m|n)$ Polychronakos spin chain [14,15]. Consequently,
by using the motif representations and skew-Young 
diagrammes associated with supersymmetric 
Polychronakos spin chain [15], one can in principle calculate
the degeneracy factors for all energy eigenvalues of supersymmetric
HS spin chain. However, similar to the nonsupersymmetric case, 
this method for finding the full spectrum and 
related partition function becomes very complicated for large values 
of $N$. 

The aim of the present article is to 
find out the exact partition function for supersymmetric $SU(m|n)$
HS model by applying the 
freezing trick and also to study global properties like level density 
distribution of the corresponding spectrum. For this purpose, 
it is convenient to map the supersymmetric HS model to a usual spin chain  
containing an `anyon like' representation of the permutation algebra
as spin dependent interactions [16,17]. 
In Sec.2 we describe this mapping and also show how the freezing trick 
can be applied for the case of $SU(m|n)$ HS spin chain by embedding it in a  
spin CS model containing the same 
anyon like representation of the permutation algebra. In Sec.3,  we 
find out the complete spectrum of such spin CS model
including the degeneracy factors for all energy levels. In Sec.4,
we calculate the partition function of this spin CS model 
at the strong coupling limit and divide it by that of the spinless
CS model to finally obtain 
the partition function of $SU(m|n)$ HS spin chain. 
In this section, we also discuss about the motif representation for 
$SU(m|n)$ HS spin chain and find that, due to the lifting 
of a selection rule, some extra energy levels appear in the 
spectrum in comparison with the case of $SU(m)$ spin chain. 
 Subsequently, we conjecture that the partition function of 
$SU(m|n)$ HS model satisfies a duality relation under the 
exchange of bosonic and fermionic spin degrees of freedom. 
In Sec.5, we study the level density distribution
and the nearest neighbour spacing distribution for the spectrum of 
$SU(m|n)$ HS spin chain by using its exact partition function. 
It is found that, for sufficiently large values of $N$,  
 continuous part of the energy level density
satisfies the Gaussian distribution with a high degree of accuracy.
We also derive exact expressions for the mean value and 
standard deviation which characterize such 
Gaussian distribution. Sec.6 is the concluding section.

\noindent \section {Application of the freezing trick }
\renewcommand{\theequation}{2.{\arabic{equation}}}
\setcounter{equation}{0}

For the purpose of defining the  $SU(m|n)$ supersymmetric HS spin chain,  
let us consider a set of operators like
$C_{j \alpha}^\dagger$($C_{j \alpha}$) which creates (annihilates)
a particle of species $\alpha$ on the $j$-th lattice site.
These creation (annihilation) operators are assumed to be bosonic when 
$\alpha \in [1,2,....,m]$ and fermionic when $\alpha \in [m+1,m+2,....,m+n]$.
Thus,  the parity of $C_{j \alpha}^\dagger$($C_{j \alpha}$) is defined as 
\bea
 &&p(C_{j \alpha})=p(C_{j \alpha}^\dagger)=0 ~
\mathnormal{for}~ \alpha \in [1,2,....,m] \, , \nn \\
 &&p(C_{j \alpha})=p(C_{j \alpha}^\dagger)=1 ~
 \mathnormal{for}~ \alpha \in [m+1,m+2,....,m+n] \, . \nn
\eea
These operators satisfy commutation (anti-commutation) relations like 
\beq
[C_{j \alpha},C_{k \beta}]_{\pm}=0 \, ,~ 
[C_{j \alpha}^\dagger,C_{k \beta}^\dagger]_{\pm}=0 \, , ~
[C_{j \alpha},C_{k \beta}^\dagger]_{\pm}=\delta_{jk}\delta_{\alpha \beta} \, ,
\label{b1}
\eeq
where $[A,B]_{\pm} \equiv AB- (-1)^{p(A)p(B)}BA$.
Next, we focus our attention to a subspace of 
the related Fock space, for
which the total number of particles per site is always one:

\beq
 \sum_{\alpha=1}^{m+n} C_{j\alpha}^{\dagger} C_{j\alpha}=1,
\label{b2}
\eeq
for all $j$. On the above mentioned subspace, 
one can define supersymmetric exchange operators as 
\beq
\hat{P}_{jk}^{(m|n)} \equiv
\sum_{\alpha,\beta=1}^{m+n} C_{j \alpha}^\dagger
C_{k \beta}^\dagger C_{j \beta}C_{k \alpha} \, ,
\label{b3}
\eeq
where $1 \leq j <k \leq N$.  These 
$\hat{P}_{jk}^{(m|n)}$'s yield 
a realization of the permutation algebra given by
\beq
\mathcal{P}_{jk}^2=1 \, , 
~\mathcal{P}_{jk}\mathcal{P}_{kl}=\mathcal{P}_{jl}\mathcal{P}_{jk}
=\mathcal{P}_{kl}\mathcal{P}_{jl} \, ,
 ~[\mathcal{P}_{jk},\mathcal{P}_{lm}]=0 \, ,
\label{b4}
\eeq
where $j,~k,~l,~m$ are all distinct indices.
 Replacing $P_{jk}$ by $\hat{P}_{jk}^{(m|n)}$ in eqn.(\ref{a1}),
 one obtains the Hamiltonian of $SU(m|n)$ supersymmetric 
  HS model as [3]
\beq
\mathcal{H}^{(m|n)}= \frac{1}{2} \sum_{1\leq j<k\leq N} 
\frac{ \left(1+\hat{P}_{jk}^{(m|n)}\right)}{\sin^2(\xi_j-\xi_k)}.
\label{b5}
\eeq

Now we want to describe how this $SU(m|n)$ supersymmetric 
HS model (\ref{b5}), containing creation-annihilation operators,   
can be transformed to a spin chain. 
To this end, we consider a particular type of 
anyon like representation 
of permutation algebra (\ref{b4}), which acts on 
a spin state like $\v \alpha_1 \alpha_2 \dots \alpha_N \r $
 (with $\alpha_j \in [1,2,\dots,m+n]  $) as [16,17]
\beq
\tilde{P}_{jk}^{(m|n)} \v \alpha_1  \dots \alpha_j \dots 
\alpha_k \dots \alpha_N \r
= e^{i \Phi (\alpha_j,\alpha_{j+1},\dots,\alpha_k)}
\v \alpha_1  \dots \alpha_k \dots \alpha_j \dots \alpha_N \r,
\label{b6}
\eeq
where 
$e^{i \Phi(\alpha_j,\alpha_{j+1},\dots,\alpha_k)}= 1 $ 
if $\alpha_j,\alpha_k \in [1,2,\dots,m]$,
$e^{i \Phi(\alpha_j,\alpha_{j+1},\dots,\alpha_k)}= -1 $ 
if $\alpha_j,\alpha_k \in [m+1,m+2,\dots,m+n]$, and 
$e^{i \Phi(\alpha_j,\alpha_{j+1},\dots,\alpha_k)}= 
(-1)^{{\pi \sum_{p=j+1}^{k-1} \sum_{\tau=m+1}^{m+n} 
\delta_{\alpha_p, \tau }}} $ 
if $\alpha_j\in [1,2,\dots,m]$  and 
$ \alpha_k \in [m+1,m+2, \dots,m+n]$ or  vice versa.
For the purpose of interpreting the phase factor 
$e^{i \Phi(\alpha_j,\alpha_{j+1},\dots,\alpha_k)}$ in a physical way, 
it is convenient to call $ \alpha_i $  a `bosonic' spin when $ 
\alpha_i \in [1,2,\dots,m] $ and 
 a `fermionic' spin when $\alpha_i \in [m+1,m+2,\dots,m+n] $.
From eqn.(\ref {b6}) it follows that, the exchange of two bosonic (fermionic)
spins produces a phase factor of $ 1 \, (-1)$ irrespective of the nature of 
spins situated in between the $j$-th and $k$-th lattice sites. However, if
we exchange one bosonic spin with one fermionic spin,  then the phase factor
becomes $(-1)^\rho$ where $\rho$ is the total 
number of fermionic spins situated in between the
$j$-th and $k$-th lattice sites. 

Next we observe that,  due to the constraint (\ref{b2}), 
the Hilbert space associated with $SU(m|n)$ HS Hamiltonian (\ref{b5}) 
can be spanned through the following 
orthonormal basis
vectors: $ C_{1 \alpha_1}^\dagger C_{2 \alpha_2}^\dagger 
\dots C_{N \alpha_N}^\dagger \v 0 \r $, where $\v 0\r$ is the
 vacuum state and $\alpha_j \in [1,2,\dots,m+n]$. Consequently, it is 
possible to define a one-to-one mapping  between 
these basis vectors  and those of the 
 above mentioned spin chain as
\beq
\v\alpha_1 \alpha_2 \dots \alpha_N \r \leftrightarrow C_{1 \alpha_1}^\dagger
C_{2 \alpha_2}^\dagger \dots C_{N \alpha_N}^\dagger \v 0 \r.
\label{b7}
\eeq
With the help of commutation (anti-commutation) relations (\ref{b1}),  
 one can easily verify that 
\bea
&&\hat{P}_{jk}^{(m|n)} \, C_{1 \alpha_1}^\dagger 
\dots C_{j \alpha_j}^\dagger \dots
C_{k \alpha_k}^\dagger \dots C_{N \alpha_N}^\dagger \v 0 \r  \nn \\
&&~~~~~~~~~~~~~~~~~~~~~~~~~~
\, = \, e^{i\Phi(\alpha_j,\dots,\alpha_k)} \, C_{1 \alpha_1}^\dagger 
 \dots C_{j \alpha_k}^\dagger \dots C_{k \alpha_j}^\dagger 
\dots C_{N \alpha_N}^\dagger \v 0\r, ~~~~
\label{b8}
\eea
where $e^{i\Phi(\alpha_j,\dots,\alpha_k)}$ is the same phase factor 
which appeared in eqn.(\ref{b6}). 
Comparison of eqn.(\ref{b8}) with eqn.(\ref{b6}) through the mapping 
(\ref{b7}) reveals that, the anyon like representation 
 $\tilde{P}_{jk}^{(m|n)}$  is equivalent to the supersymmetric 
exchange operator $\hat{P}_{jk}^{(m|n)}$.
Hence, if we define a spin chain Hamiltonian through
 $\tilde{P}_{jk}^{(m|n)}$ as
\beq
H^{(m|n)}= \frac{1}{2} \sum_{1\leq j<k\leq N} 
\frac{ \left(1+\tilde{P}_{jk}^{(m|n)}\right)}{\sin^2(\xi_j-\xi_k)},
\label{b9}
\eeq
that would be completely equivalent to the
supersymmetric $SU(m|n)$ HS model (\ref{b5}) [16]. 
Clearly,  for the special case $n=0$, $\tilde{P}_{jk}^{(m|n)}$ reproduces 
the original spin exchange operator $P_{jk}$ and 
${H}^{(m|n)}$ (\ref{b9}) reduces to the 
Hamiltonian of $SU(m)$ HS spin chain (\ref{a1}). 
Since it is convenient to apply the freezing trick to 
the spin chain Hamiltonian (\ref {b9}), for the rest of this 
article we shall deal with  this form of 
supersymmetric $SU(m|n)$ HS model 
instead of its original form (\ref{b5}).

By using the anyon like representation
 $\tilde{P}_{jk}^{(m|n)}$, one can construct a spin CS model like 
\beq
H^* =-\sum_{j=1}^N \frac{\partial^2}{\partial x_j^2} + 
2a\sum_{1\leq j < k \leq N} \frac{ \big( \, a+\tilde{P}_{jk}^{(m|n)} \, \big) }
{\sin^2(x_j-x_k)} \, ,
\label{b10}
\eeq
which contains spin as well as dynamical degrees of freedom and 
the positive parameter $a$ as coupling constant.  
With the help of mapping (\ref {b7}) it can be shown that, 
this spin CS model is equivalent to a supersymmetric 
spin CS model [18] with $Y(gl_{(m|n)})$ super-Yangian symmetry.  
The spin CS Hamiltonian $H^*$ (\ref {b10}) might be formally written as 
\beq
H^*=H_0+4a \mathsf{H}^{(m|n)},
\label{b11}
\eeq
where $H_0$ is the Hamiltonian of spinless CS model given by [19] 
\beq
H_0=-\sum_{j=1}^N \frac{\partial^2}{\partial x_j^2} + 
2a(a-1)\sum_{1\leq j < k \leq N} \frac{1}{\sin^2(x_j-x_k)} \, ,
\label{b12}
\eeq
and 
$\mathsf{H}^{(m|n)}$ is obtained from $H^{(m|n)}$ (\ref{b9})
 by the replacement
$ \xi_j \rightarrow x_j$.
Now the decoupling of the dynamical degrees of freedom of $H^*$ (\ref {b10})
from its spin degrees of freedom can be achieved by using 
the freezing trick [10-12]. This trick is based
on the fact that in the limit $a \rightarrow \infty $,
particles freeze at the
equilibrium positions of $H_0$, which are simply the lattice points ($\xi_j$)
of the spin chain in eqn.(\ref{b9}). 
Consequently,  by using eqn.(\ref{b11})
at freezing limit, we find that the energy levels of $H^*$
are approximately given by 
\beq
E^*_{jk}\simeq E_{0,j}+4aE_k,
\label{b13}
\eeq
where $E_{0,j}$ and $E_k$ are any two levels of $H_0$  
and $H^{(m|n)}$ respectively. 
Hence, we obtain a relation like 
\beq
Z^{(m|n)}(T)=\lim_{a\rightarrow \infty} \frac{Z^*(4aT)}{Z_0(4aT)},
\label{b14}
\eeq
where $Z^{(m|n)}$, $ Z^*$ and $Z_0$ denote the partition functions 
corresponding to the Hamiltonians
$H^{(m|n)}$,  $H^*$ and $H_0$ respectively. Thus the
 freezing trick allows us to compute the partition function of
 $SU(m|n)$ supersymmetric HS spin chain, by modding out   
 the contribution of spinless CS model from the partition function of
 spin CS model (\ref{b10}).  Due to the 
Gallelian invariance of $H^*$ and $H_0$ it follows that, if 
$\psi$ is an eigenstate of any one of these Hamiltonians with momentum $p$, 
then $\psi'=e^{2i\tau \sum_{j=1}^N x_j} \psi$ will also be an 
eigenstate of the same Hamiltonian with momentum $(p+2 \tau N)$.
As a result, we can always adjust the 
parameter $\tau$ such that $\psi'$ will be an eigenfunction of
 $H^*$ or $H_0$ with zero momentum.
In this article, we shall always 
consider eigenstates of these Hamiltonians with zero momenta and evaluate the 
partition functions $Z^*$ as well as $Z_0$ at the center of mass 
frame. Since both $Z^*$ and $Z_0$ get 
modified by the same multiplicative factor due to a Gallelian transformation, 
$Z^{(m|n)}$ does not depend on the choice of the reference frame.

\noindent \section {Spectrum of spin CS model }
\renewcommand{\theequation}{3.{\arabic{equation}}}
\setcounter{equation}{0}
In this section our aim is to find out the complete spectrum of 
spin CS model (\ref{b10})
containing anyon like representation of the permutation algebra. 
Even though the spectrum of such spin CS model has been studied
earlier [17],   multiplicities of degenerate eigenfunctions 
corresponding to all energy levels 
have not been found. Since these numbers are required 
for calculating the partition function of this model, 
here we want to derive a general expression  
 for the degeneracy factors of all energy levels. 
It is well known that the eigenfunctions of  spin 
CS Hamiltonian (\ref {b10}) can be written in a factorised form like 
\beq
\psi (x_1,\dots,x_N;\alpha_1,\dots,\alpha_N)
=\Gamma^a \phi(x_1,\dots,x_N ; \alpha_1,\dots,\alpha_2),
\label{c1}
\eeq
where $\Gamma=\prod_{i<j} \sin(x_i-x_j)$.
 By operating $H^*$ (\ref {b10}) on the above form of $\psi$, 
 we find that
\beq
H^*\psi=\Gamma^a \tilde{H}^* \phi,
\label{c2}
\eeq
where
\bea
\tilde{H}^*=4\biggl[ \sum_j \biggl( z_j \frac{\partial}
{\partial z_j}\biggr) ^2 + 
a \sum_{k<j} \frac{z_k+z_j}{z_k-z_j} 
\biggl(z_k\frac{\partial}{\partial z_k}-z_j\frac{\partial}
{\partial z_j}\biggr) ~~~~~~~~~~\nn \\
~~~~~~~~-2a \sum_{k<j} (1+\tilde{P}^{(m|n)}_{jk}) \frac{z_jz_k}
{(z_j-z_k)^2} + \frac{a^2}{12} N(N^2-1)\biggr],
\label{c3}
\eea
with $z_j=e^{2ix_j}$. Equation (\ref {c2}) implies 
that,  if $\phi$ is an eigenvector of $\tilde{H}^*$ with 
 eigenvalue $E$, then 
$\Gamma^a \phi$  would be an eigenvector of $H^*$ with the 
same eigenvalue. Thus the diagonalisation problem of 
 $H^*$ is reduced to the diagonalisation problem of $\tilde{H}^*$.

For solving $\tilde{H}^*$, it is convenient to introduce 
another operator $\mathcal{H}$ 
which acts only on the coordinate degree of freedom and
 may be given by [7]
\bea
\mathcal{H}=4\biggl[ \sum_j \biggl( z_j \frac{\partial}
{\partial z_j}\biggr) ^2 + a \sum_{k<j} 
\frac{z_k+z_j}{z_k-z_j} \biggl(z_k\frac{\partial}
{\partial z_k}-z_j\frac{\partial}{\partial z_j}\biggr) ~~~~~~~~~~ \nn \\
~~~~~~~~-2a \sum_{k<j} (1- K_{jk}) \frac{z_jz_k}{(z_j-z_k)^2} 
+ \frac{a^2}{12} N(N^2-1)\biggr],
\label{c4}
\eea
where the $K_{jk}$ is the coordinate exchange operator 
which exchanges the coordinates of $j$-th and $k$-th particle:
\beq
K_{jk}f(x_1,\dots,x_j,\dots,x_k,\dots,x_N)=
f(x_1,\dots,x_k,\dots,x_j,\dots,x_N).\nn
\eeq
It may be observed that,  $\tilde{H}^*$ (\ref{c3}) 
can be reproduced from the expression of 
$\mathcal{H}$  (\ref{c4}) through the substitution
$K_{jk}\rightarrow -\tilde{P}_{jk}^{(m|n)}$. This 
connection between $\tilde{H}^*$ and $\mathcal{H}$ will play a 
crucial role in our calculation for finding the spectrum of $\tilde{H}^*$. 
Let us first consider state vectors given by monomials like
\beq
\xi_{\bf{p}}=z_1^{p_1}z_2^{p_2}\dots z_N^{p_N},
\label{c5}
\eeq
where ${\bf p} \equiv \{p_1,p_2,\dots,p_N\} \in \mathbb{R}^N$
 satisfies the constraints: 
(i) $(p_i-p_j)$ are integers for all $i,j$ and 
(ii) $\sum_{i=1}^N p_i=0$. The last condition implies that 
these monomials represent state vectors with zero total momentum.  
In particular, one can consider 
$\xi_{\bf\hat{p}}$
corresponding to a nonincreasing vector
${\bf\hat{p}} \equiv (\hat{p}_1,\hat{p}_2,\dots,\hat{p}_N)$,  
whose elements satisfy the conditions: 
(i) $l_i\equiv \hat{p}_{i}-\hat{p}_{i+1}$ is a nonnegative integer
for $i \in [1,\dots,N-1]$ and 
(ii) $\sum_{i=1}^N {\hat p}_i=0$. 
It is evident that, 
$N-1$ number of nonnegative integers ($l_i$'s) are sufficient to specify  
a nonincreasing vector ${\bf\hat{p}}$. 
Given two distinct nonincreasing vectors 
${\bf\hat{p}}$ and 
${\bf\hat{p}'}$, we shall write ${\bf\hat{p}} \prec {\bf\hat{p}'}$ 
if $\hat{p}_1-\hat{p}_1'=\dots=\hat{p}_{i-1}-\hat{p}_{i-1}'=0$ 
and $\hat{p}_i<\hat{p}_i'$.
A partial ordering can be defined on monomials like 
$\xi_{\bf{p}}$ (\ref {c5}) in the following way. 
By permuting the elements of any  ${\bf p}$, 
one can always construct a  unique nonincreasing vector 
${\bf\hat{p}}$.  
The basis element $\xi_{\bf p}$ would precede  $\xi_{\bf p'}$ 
if $ {\bf \hat{p}} \prec {\bf \hat{p}'} $, where 
$ {\bf \hat{p}}$ and $ {\bf \hat{p}'}$ are 
 nonincreasing vectors obtained
 from ${\bf p}$ and ${\bf p'}$ respectively by permuting 
their components. The above defined ordering is effectively a 
partial ordering, since it does not induce an
ordering between $\xi_{\bf p}$ and $\xi_{\bf p'}$ 
when ${\bf \hat{p}}= {\bf\hat{p}'}$. It can be shown 
that the action of $\mathcal{H}$ (\ref {c4}) on the state  
vector $\xi_{\bf p}$ yields [7,9]
\beq
\mathcal{H } \xi_{\bf p} \, = \,  E_{\bf \hat{p}}  \xi_{\bf p} + 
\sum_{\substack{{\bf p'}\\({\bf \hat{p}'}<
{\bf \hat{p}})}}c_{{\bf p} {\bf p'}} \xi_{\bf p'} \, ,
\label{c6}
\eeq
where
\beq
 E_{\bf \hat{p}}=\sum_{i=1}^N \left \{2 \hat{p_i}+a(N+1-2i) \right \}^2.
\label{c7}
\eeq
Thus it is clear that, if one constructs a Hilbert
space through basis vectors of the form (\ref{c5}) and 
partially order them in the
above mentioned way, then $\mathcal{H}$ will act as a 
triangular matrix on this space.

Next we want to construct another partially ordered Hilbert space,
on which $\tilde{H}^*$ (\ref{c3}) can be represented as a triangular matrix.
To this end, we define a set of permutation operators as 
$\Pi_{jk}^{(m|n)}=\tilde{P}_{jk}^{(m|n)}K_{jk}$.
Since both  $\tilde{P}_{jk}^{(m|n)}$ and $K_{jk}$ 
satisfy an algebra of the form (\ref{b4}), while acting 
on the spin and coordinate spaces respectively, the newly defined 
operator $\Pi_{jk}^{(m|n)}$ also yields 
a representation of the same permutation algebra 
on the direct product of coordinate and spin spaces.
Hence, by using this representation of permutation 
algebra, we can construct a `generalized'
antisymmetric projection operator $\Lambda^{(m|n)} $ 
satisfying the relation 
\beq
\Pi_{jk}^{(m|n)}\Lambda^{(m|n)}=
\Lambda^{(m|n)} \Pi_{jk}^{(m|n)} =-\Lambda^{(m|n)} \, ,
\label{c8}
\eeq
or, equivalently, $\tilde{P}_{jk}^{(m|n)} 
\Lambda^{(m|n)}= -K_{jk}\Lambda^{(m|n)}$ [16,17]. 
Even though  
 $\Lambda^{(m|n)} $ can be expressed as a function of 
$\Pi_{jk}^{(m|n)}$,  explicit form of this projection operator
is not necessary for our present purpose. However it may be noted that, 
since  both $K_{jk}$ and $\tilde{P}_{jk}^{(m|n)}$ 
commute with $\mathcal{H}$ (\ref {c4}), 
$\Lambda^{(m|n)}$ also satisfies the relation 
\beq
[\mathcal{H},\Lambda^{(m|n)}]=0 \, .
\label{c9}
\eeq
With the help of projection operator $\Lambda^{(m|n)}$, 
we define a state vector on the direct product of coordinate and 
spin spaces as 
\bea
\phi^{\boldsymbol{\alpha}}_{\bf p} \equiv 
\phi^{\alpha_1 \dots \alpha_N}_{p_1 \dots p_N}= \Lambda^{(m|n)} 
\{ \xi_{\bf p} \v \alpha_1 \dots \alpha_N \r \} .
\label{c10}
\eea
Using  eqns.(\ref{c8}) and (\ref{b6}) it can be shown that
\bea
\phi^{\alpha_1\dots\alpha_j\dots\alpha_k \dots 
\alpha_N}_{ p_1 \dots \,  p_j\dots \,  p_k \dots \, p_N} 
\hskip -.56 cm  
&&=-\Lambda^{(m|n)} 
 K_{jk} {\tilde P}_{jk}^{(m|n)}
 \{ z_1^{p_1} \dots z_j^{p_j} \dots z_k^{p_k} \dots z_N^{p_N} 
\v \alpha_1 \dots\alpha_j \dots \alpha_k \dots \alpha_N \r \} \nn \\
&&= -e^{i \Phi(\alpha_j,\dots,\alpha_k)}\Lambda^{(m|n)}  
 \{ z_1^{p_1} \dots z_j^{p_k} \dots z_k^{p_j} \dots z_N^{p_N} 
\v \alpha_1 \dots \alpha_k \dots \alpha_j \dots \alpha_N  \r \} \nn \\
&&= -e^{i \Phi(\alpha_j,\dots,\alpha_k)} \phi^{\alpha_1 \dots 
\alpha_k\dots \alpha_j \dots \alpha_N}_{p_1 \dots \, p_k \dots
\, p_j \dots \, p_N} \, .
\label{c11}
\eea
By repeatedly using the above equation we find that 
\beq
\phi_{\bf p}^{\boldsymbol{\alpha}}= \epsilon ({\boldsymbol{\alpha}}, {\bf p})
\, \phi_{\bf \hat{p}}^{\boldsymbol{\alpha'}},
\label{c12}
\eeq
where $ \epsilon ({\boldsymbol{\alpha}}, {\bf p}) = \pm 1$, ${\bf \hat p}$
is the nonincreasing vector corresponding to ${\bf p}$ and 
${\boldsymbol{\alpha'}} $ is a spin vector which
is obtained by permuting the components of ${\boldsymbol{\alpha}}$. 
Hence, all state vectors of the form (\ref {c10}) can be obtained 
by choosing ${\bf p}$ from the set of nonincreasing vectors only. 

Corresponding to any given nonincreasing vector ${\bf \hat{p}}$, 
one can define a vector space as
\beq
\mathbb{V}_{\bf \hat{p}} \equiv 
\l \, \phi_{\bf \hat{p}}^{\boldsymbol{\alpha}}~~|
\, \alpha_1 , \dots ,\alpha_N \in [1,2, \dots m+n] \, \r \,  . 
\label{c13}
\eeq
It is important to note that,  different values of 
${\boldsymbol{\alpha}}$ may lead to the same 
$\phi_{\bf \hat{p}}^{\boldsymbol{\alpha}}$ which is a 
basis element of $\mathbb{V}_{\bf \hat{p}} $.
 To see this thing in a simple way, let us take a nonincreasing
 sequence ${ \bf \hat{p}}$ satisfying the condition  
$ \hat{p}_i = \hat{p}_j = \hat{p}$ (say). 
For this special case, eqn.(\ref{c11}) reduces to 
\beq
\phi^{\alpha_1 \dots \alpha_i \dots \alpha_j \dots \alpha_N}_{ \,
\hat{p}_1 \,  \dots \, \hat{p} \, \dots \, \hat{p} \, \dots \,  \hat{p}_N }
 = -\, e^{i \Phi (\alpha_i,\dots ,\alpha_j)} \,
\phi^{\alpha_1 \dots \alpha_j \dots \alpha_i \dots \alpha_N}_{ \, 
\hat{p}_1 \,  \dots \, \hat{p} \, \dots \, \hat{p} \, \dots \,  \hat{p}_N }\, .
\label{c14}
\eeq
Clearly 
$\phi^{\alpha_1 \dots \alpha_i \dots \alpha_j \dots \alpha_N}_{ \, 
\hat{p}_1 \,  \dots \, \hat{p} \, \dots \, \hat{p} \, \dots \,  \hat{p}_N }$
and $\phi^{\alpha_1 \dots \alpha_j \dots \alpha_i \dots \alpha_N}_{ \,  
\hat{p}_1 \,  \dots \, \hat{p} \, \dots \, \hat{p} \, \dots \,  \hat{p}_N }$
represent the same state vector (up to a phase factor), 
 although they correspond to different values of ${\boldsymbol{\alpha}}$.
For a given $\phi_{\bf \hat{p}}^{\boldsymbol{\alpha}}$, 
we say that two spin components of 
${\boldsymbol{\alpha}}$ belong to the same `sector' if 
the corresponding two components of ${ \bf \hat{p}}$ are equal to 
each other. For example, the spin components 
$\alpha_i$ and $ \alpha_j$ appearing in the state
$\phi^{\alpha_1 \dots \alpha_i \dots \alpha_j \dots \alpha_N}_{ \, 
\hat{p}_1 \,  \dots \, \hat{p} \, \dots \, \hat{p} \, \dots \,  \hat{p}_N }$ 
belong to the same sector according to this convention. 
Since $e^{i \Phi}=1$, for $\alpha_i,\alpha_j 
\in [1,2,\dots,m]$, it is clear from eqn.(\ref{c14})
that bosonic spins within the same sector 
obey `fermionic statistics' after antisymmetrisation. In particular, 
two bosonic spins of same flavour can not coexist within a single sector. 
Similarly, since $e^{i \Phi}=-1$ for $\alpha_i,
\alpha_j \in [m+1,m+2,\dots,m+n]$,
one can find from eqn.(\ref{c14}) that fermionic spins within the 
same sector obey `bosonic statistics' after antisymmetrisation.
Therefore, any number of fermionic spins having the same flavour can
be accommodated within a single sector. 

Now we want to find out the dimensionality of the space 
$\mathbb{V}_{\bf \hat{p}}$. For this purpose, it is useful to write
${\bf \hat{p}}$ in the form 
\beq
\bf{\hat{p}} \equiv \big ( \overbrace{\rho_1,\dots,\rho_1}^{k_1},
\dots,\overbrace{\rho_i,\dots,\rho_i}^{k_i},
\dots,\overbrace{\rho_r,\dots,\rho_r}^{k_r} \big ),
\label{c15}
\eeq 
where $\rho_1>\dots>\rho_i>\dots>\rho_r$, $\sum_{i=1}^r k_i=N$, 
and $r$ is an integer which can take any value from $1$ to 
$N$. It is obvious that  ${\bf k} \equiv \{ k_1,\dots,k_r\}$, which  
belongs to the set $\mathcal{P}_N$ of ordered partitions of $N$, 
 may be treated as a function of 
$\bf{\hat{p}}$. For a given $\phi_{\bf \hat{p}}^{\boldsymbol{\alpha}}$,
 clearly the components of $\boldsymbol{\alpha}$ are 
 separated into $r$ different sectors where the 
$i$-th sector contains $k_i$ number of spins. 
It is evident that the dimensionality of the space 
$\mathbb{V}_{\bf \hat{p}}$ may be obtained by counting 
the number of independent ways one can distribute total 
$N$ number of spins within $r$ sectors. To this end, let us first 
try to find out the number of independent ways of filling up the $i$-th sector 
through $j_1$ number of bosonic spins and $j_2$ number of fermionic  
spins, where $j_1+j_2=k_i$. Using eqn.(\ref{c14}) we have already seen that, 
bosonic and fermionic spins within the same sector obey fermionic and 
bosonic statistics respectively. Therefore, we can pick up 
$j_1$ number of bosonic spins
from $m$ different flavours in $^m C_{j_1}$ different 
ways and $j_2$ number of fermionic spins
from $n$ different flavours in $^{j_2+n-1}C_{j_2}$ ways,  where $^p 
C_l=\frac{p\, !}{l\hskip .01 cm ! \, (p-l)!} $  for $l \leq p$ and 
 $^{p}C_l=0$ for $l>p $.  
Thus the number of independent ways of 
filling up the $i$-th sector through $j_1$ number of bosonic spins and 
$j_2$ number of fermionic spins is given by 
$$
^mC_{j_1} \, ^{n+j_2-1}C_{j_2} \, .
$$
Summing up these numbers for all possible values of 
$j_1$ and $j_2$, we obtain the total number of independent ways 
of filling up the $i$-th sector through  $k_i$ number of spins 
as
\bea
d^{(m|n)}(k_i)  
&=& \sum_{j_1= ~0}^{k_i} 
 \big(\,  ^mC_{j_1} \, ^{n+k_i-j_1-1}C_{k_i-j_1} \, \big)  \nn \\
&=& \sum_{j_1= ~0}^{{\rm min} \, (m,k_i)} 
\frac{m!~(n+k_i -j_1 -1)!}{j_1! ~(m-j_1)!~(n-1)!~(k_i-j_1)!}\, .
\label{c16}
\eea
Since two spins belonging to different sectors
do not follow any exchange relation like (\ref {c14}),
the number of independent ways we can distribute total 
$N$ number of spins within $r$ different sectors is
given by the product of all $d^{(m|n)}(k_i)$. 
Therefore, by using  (\ref{c16}), we finally obtain 
the dimension of $\mathbb{V}_{\bf \hat{p}}$ as 
\beq
d^{(m|n)}({\bf k})=\prod_{i=1}^r 
d^{(m|n)}(k_i)=\prod_{i=1}^r 
 \left( \sum_{j=0}^{{\rm min}\, (m,k_i)} \,
   {}^mC_j \, {}^{n+k_i-j-1}C_{k_i-j}  \right)  \, .
\label{c17}
\eeq
Even though this expression is derived by assuming that bosonic and 
fermionic spin degrees of freedom (i.e., $m$ and $n$ respectively) take 
nonzero values, it is also possible 
to obtain the dimension of $\mathbb{V}_{\bf \hat{p}}$ for the
$SU(n)$ fermionic case by putting $m=0$ in eqn.(\ref{c17}):
\beq
d^{(0|n)}({\bf k})= \prod_{i=1}^r 
~^{n+k_i-1}C_{k_i} \, .
\label{c18}
\eeq
Furthermore, by putting $n=0$ in 
eqn.(\ref{c17}), subsequently using the relation $^{p}C_l=0$ for $l>p\geq 0$,  
and also assuming that $^{-1}C_0=1$, 
one can reproduce the
dimension of $\mathbb{V}_{\bf \hat{p}}$ for the $SU(m)$ 
bosonic case [9] as
\beq
d^{(m|0)}({\bf k})= \prod_{i=1}^r 
~^{m} C_{k_i} .
\label{c19}
\eeq
It is interesting to observe that, while 
$d^{(m|0)}({\bf k})$ (\ref {c19}) can take a nonzero value 
only if $k_i \leq m $ for all $i$, both 
$d^{(m|n)}({\bf k})$ (\ref{c17}) and 
$d^{(0|n)}({\bf k})$ (\ref{c18}) take nonzero values 
for any ${\bf k} \in \mathcal{P}_N$. Consequently, 
$\mathbb{V}_{\bf \hat{p}}$ will represent a nontrivial vector space
for the $SU(m)$ bosonic case only if at most $m$ components of 
${\bf \hat{p}}$ take the same value. On the other hand, 
$\mathbb{V}_{\bf \hat{p}}$ will represent a nontrivial vector space
for all possible values of ${\bf \hat{p}}$ when at least one 
fermionic spin degrees of freedom is present.  

The Hilbert space associated with 
$\tilde{H}^*$ (\ref{c3}) may now be defined by taking the direct sum of
 $\mathbb{V}_{\bf \hat{p}}$ (\ref {c13})  for all allowed values of 
${\bf \hat{p}}\, $: 
\beq 
\mathbb{V} 
= \mathop{\oplus}_{\bf\hat{p}} \mathbb{V}_{\bf \hat{p}} \, .
\label{c20}
\eeq
We define a partial ordering in this Hilbert space by saying that 
the basis element 
$\phi_{\bf \hat{p}}^{\boldsymbol{\alpha}}$
precedes  
$\phi_{\bf \hat{p'}}^{\boldsymbol{\alpha'}}$ if 
$ {\bf \hat{p}} \prec {\bf \hat{p}'} $. 
By consecutively applying the relations (\ref{c10}), (\ref{c8}),  
(\ref{c9}), (\ref{c6}) and (\ref{c12}), it is easy to check that
\bea
\tilde{H}^* (\phi^{\boldsymbol{\alpha}}_{\bf \hat{p}})
&=&\Lambda^{(m|n)} \, \mathcal{H} \, \xi_{\bf \hat{p}} 
\v \alpha_1 \dots \alpha_N \r \nn \\
&=&\Lambda^{(m|n)} \biggl( E_{\bf \hat{p}}  \xi_{\bf \hat{p}} + 
\sum_{\substack{{\bf p'}\\({\bf \hat{p}'}
<{\bf \hat{p}})}}c_{{\bf \hat{p}}
 {\bf p'}} \,  \xi_{\bf p'}\biggl) \v \alpha_1 
\dots \alpha_N \r \nn \\
&=&E_{\bf \hat{p}}  
\phi_{\bf \hat{p}}^{\boldsymbol{\alpha}} \,  + \, 
\sum_{\substack{{\bf p'}\\({\bf \hat{p}'}
<{\bf \hat{p}})}}
\epsilon ({\boldsymbol{\alpha}}, {\bf p'}) \,
c_{{\bf \hat{p}} {\bf p'}} \,
\phi_{\bf \hat{p}'}^{\boldsymbol{\alpha'}} \, .
\label{c21}
\eea
Hence  $\tilde{H}^*$ is represented as a triangular matrix on  
$\mathbb{V}$. Diagonal elements of this triangular matrix yield
the eigenvalues of $\tilde{H}^*$ as  
\beq
E^*({\bf \hat{p}},\boldsymbol{\alpha}) \equiv 
E_{\bf \hat{p}}= \sum_{i=1}^N \big\{  2 \hat{p}_i +a(N+1-2i)  \big\}^2.
\label{c22}
\eeq
Consequently, the eigenvalues of spin CS Hamiltonian 
$H^*$ (\ref {b10}) are also given by 
$E^*({\bf \hat{p}},\boldsymbol{\alpha})$ in the above equation.

Since $E^*({\bf \hat{p}},\boldsymbol{\alpha})$ in eqn.(\ref{c22})
does not really depend on the spin vector $\boldsymbol{\alpha}$, 
the number of degenerate energy eigenstates associated with the
 quantum number ${\bf \hat{p}}$ would coincide with the dimension of the 
space $\mathbb{V}_{\bf \hat{p}} $. Thus the degeneracy factor 
of the energy eigenvalue 
 $E^*({\bf \hat{p}},\boldsymbol{\alpha})$ corresponding to the 
 quantum number ${\bf \hat{p}}$ is given by 
 $d^{(m|n)}({\bf k})$ appearing in eqn.(\ref{c17}). 
We have already seen that, in contrast to the pure bosonic case, 
 $d^{(m|n)}({\bf k})$ takes nonzero values
 for all possible ${\bf k} \in \mathcal{P}_N$
when at least one fermionic spin degrees of freedom is present. 
Consequently, the presence of fermionic spin degrees of freedom in 
 $H^*$ (\ref {b10}) would lead to a spectrum with 
many additional energy levels in comparison with
the spectrum of bosonic spin CS model. 

Finally let us briefly comment about the known spectrum
of spinless CS Hamiltonian $H_0$ (\ref {b12}) [19].
Using the fact that the eigenfunctions of $H_0$ 
can be written in a factorised form like
$ \psi_0=\Gamma^a \phi_0 (x_1,\dots,x_N)$,
it is possible to transform $H_0$ 
into $\tilde{H}_0$ as
$$
H_0 \psi_0=\Gamma^a \tilde{H}_0 \phi_0 \, ,
$$
where $\tilde{H}_0$ can be obtained from
$\mathcal{H}$ (\ref{c4}) through
the substitution $K_{ij} \rightarrow 1$.
For constructing the Hilbert space associated with 
$\tilde{H}_0$, one may consider elements like 
$ \phi_{\bf p} \equiv \Lambda_0 (\xi_{\bf p})$, 
where $ \Lambda_0$ is the symmetriser
in the coordinate space: $K_{jk}\Lambda_0 =  \Lambda_0$.
Since $\phi_{\bf p}=\phi_{\bf \hat{p}}$, where ${\bf \hat{p}}$ 
is the nonincreasing vector corresponding to 
${\bf p}$, the Hilbert space of $\tilde{H}_0$ is defined 
through independent basis vectors $\phi_{\bf \hat{p}}$ for 
all values of ${\bf \hat{p}}$. 
An ordering can be defined among these state vectors by saying that 
$\phi_{\bf \hat{p}}$ precedes  
$\phi_{\bf \hat{p}'}$ if $ {\bf \hat{p}} \prec {\bf \hat{p}'} $. 
Using eqn.(\ref{c6}) it can be shown that, 
$\tilde{H}_0$ acts as a triangular matrix on these completely ordered 
basis vectors and the eigenvalues of $H_0$ are also 
given by $E_{\bf \hat{p}}$ in eqn.(\ref{c22}). However, 
due to the absence of spin degrees of freedom, only one energy eigenstate 
is obtained corresponding to each quantum number ${\bf {\hat p}}$
in this case.

\noindent \section {Partition function of $SU(m|n)$ HS spin chain }
\renewcommand{\theequation}{4.{\arabic{equation}}}
\setcounter{equation}{0}

By using the freezing trick we have seen that, 
the partition function of supersymmetric $SU(m|n)$ HS spin chain
can be obtained by dividing the partition function of spin CS model (\ref {b10}) 
at the strong coupling limit through that of the spinless CS model 
(\ref {b12}).  To execute this programme, 
let us first briefly recapitulate the calculation for the partition 
function of spinless CS model (\ref{b12}) at $a \rightarrow \infty$ limit [9].
It should be noted that,  the eigenvalues in eqn.(\ref{c22}) 
 can be expanded in powers of $a$ as
\beq
E^*({\bf \hat{p}},\boldsymbol{\alpha}) \equiv 
E({\bf \hat{p}})=a^2E_0+4a \sum_{i=1}^N (N+1-2i)\hat{p}_i +O(1),
\label{d1}
\eeq
where $ E_0=\frac{1}{3} N(N^2-1) $ .
Since $E_0$ does not depend on ${\bf\hat{p}}$ or 
$\boldsymbol{\alpha}$, the effect of this $E_0$  will be manifested as the same
overall multiplicative factor in the partition functions of 
spin CS model and its spinless counterpart. 
Hence, by dropping the first term
in eqn.(\ref{d1}), and neglecting the $O(1)$ term in the limit
$a \rightarrow \infty $, one can write down 
the partition function of spinless CS model (\ref{b12}) as
\beq
Z_0(4aT) \simeq \sum_{{\bf \hat{p}} } q^{\sum_i \hat{p}_i(N+1-2i)} \, ,
\label{d2}
\eeq
where $q=e^{-1/(k_BT)}$. Using $N-1$ number of nonnegative integers
($l_i$'s) which uniquely determine ${{\bf\hat{p} }}$, 
one can further simplify this partition function as [9]
\beq
Z_0(4aT)  \simeq \sum_{l_1, \dots , l_{N-1} 
\geq 0} \prod_{j=1}^{N-1} q^{j(N-j)l_j}
 = \prod_{j=1}^{N-1}\frac{1}{ 1- q^{j(N-j)}}.
\label{d3}
\eeq

Next, we want to calculate the partition function 
of spin CS Hamiltonian (\ref{b10}) at $a \rightarrow \infty $ limit.
Dropping again the first term as well as $O(1)$ term 
from the right hand side of expansion (\ref{d1}), 
and expressing the nonincreasing vector 
${\bf \hat{p}}$ through eqn.(\ref{c15}), 
$E^*({\bf \hat{p}},\boldsymbol{\alpha})$ can be written as 
\beq
E^*({\bf \hat{p}},\boldsymbol{\alpha}) \simeq 4a \sum_{i=1}^r
 \rho_i \sum _{j=K_{i-1}+1}^{K_i} (N+1-2j),
\label{d4}
\eeq
where $K_i=\sum_{j=1}^i k_j$ denote the partial 
sums corresponding to the partition 
${\bf k} \in \mathcal{P}_N$ and $K_0=0$.
Using a set of 
variables like $\nu_j \equiv \rho_j-\rho_{j+1}$ for
$j \in [ 1,\dots,r-1 ] $ (since $\rho_j > \rho_{j+1}$,
all $\nu_j$'s are positive integers), one can express the
energy eigenvalue in eqn.(\ref{d4}) as 
\beq
E^*({\bf \hat{p}},\boldsymbol{\alpha})
 \simeq 4a\sum_{j=1}^{r-1} \nu_j N_j,
\label{d5}
\eeq
where $N_j=K_j(N-K_j)$. It may be noted that, 
due to the condition $\sum_{i=1}^N \hat{p}_i =0$,  $r-1$ number of
$\nu_j$'s uniquely determine the nonincreasing vector
${\bf \hat{p}}$  in eqn.(\ref{c15}).  Consequently, the single sum 
$\sum_{\bf\hat{p}}$  can be replaced by the double sum 
$\sum_{\mathbf{k} \in ~\mathcal{P}_N} 
\sum_{\nu_1,\dots,\nu_{r-1}>0}$ in the expression of the partition 
function.  By using the eigenvalue relation (\ref{d5}) and the  
corresponding degeneracy factor $d^{(m|n)}({\bf k})$ (\ref {c17}), 
we obtain the partition function 
of spin CS Hamiltonian (\ref{b10}) at $a \rightarrow \infty $ limit as
\bea
Z^*(4aT)& \simeq & \sum _{\bf\hat{p}}
d^{(m|n)}({\bf k})
 \prod_{j=1}^{r-1} q^{N_j \nu_j} \nn \\
&=& \sum _{\mathbf{k} \in 
~\mathcal{P}_N} d^{(m|n)}({\bf k})
\sum_{\nu_1,\dots,\nu_{r-1}>0}
 \prod_{j=1}^{r-1} q^{N_j \nu_j} \nn \\
&=&\sum _{\mathbf{k} \in 
~\mathcal{P}_N} d^{(m|n)}({\bf k}) 
\prod_{j=1}^{r-1} \frac{q^{N_j}}{1-q^{N_{j}}} ~ .
\label{d6}
\eea
Using eqns.(\ref{b14}), 
 (\ref{d3}) and (\ref{d6}), we finally obtain the partition function of 
$SU(m|n)$ HS spin chain as
\beq
Z^{(m|n)}(T)=\prod_{l=1}^{N-1} \left(1- q^{l(N-l)} \right)
\sum _{\mathbf{k} \in 
~\mathcal{P}_N} d^{(m|n)}({\bf k}) 
\prod_{j=1}^{r-1} \frac{q^{N_j}}{(1-q^{N_{j}})}.
\label{d7}
\eeq
Since the partial sums $ K_1, K_2,  \dots , K_{r-1}$ 
associated with  ${\bf k}$
are natural numbers obeying $1\leq K_1< \dots<K_{r-1}\leq N-1$, 
one can define their complements ($K_i' \,$'s) as elements of the set: 
$ \{1,\dots,N-1\}-\{K_1,\dots ,K_{r-1}\}$,  
which satisfy the ordering $ K'_1<\dots<K'_{N-r}$.
Hence one can rearrange the product
 $\prod_{l=1}^{N-1} (1-q^{l(N-l)})$ into two terms as [9]
\beq
\prod_{l=1}^{N-1} (1-q^{l(N-l)})= 
\prod_{j=1}^{r-1} (1-q^{N_j})\prod_{i=1}^{N-r}(1-q^{N_i'}),
\label{d8}
\eeq
where $N_i'=K_i'(N-K_i')$. By substituting this relation to eqn.(\ref{d7}),
we get a simplified expression for the
partition function of $SU(m|n)$ HS model as
\beq
Z^{(m|n)}(T)=\sum _{\mathbf{k}
\in ~\mathcal{P}_N} d^{(m|n)} ({\bf k})~q^{\sum
\limits^{r-1}_{j=1} N_j}\prod_{i=1}^{N-r}(1-q^{N'_{i}}) \, .
\label{d9}
\eeq
We have already seen that,  both 
$d^{(m|n)}({\bf k})$ (\ref {c17}) and 
$d^{(0|n)}({\bf k})$ (\ref {c18}) take nonzero values 
for any $\bf{k} \in \mathcal{P}_N$.
Consequently, in contrast to the restricted choice of 
$\bf{k}$ for the case of bosonic spin chain [9], 
all possible $\bf{k} \in \mathcal{P}_N$
will contribute to the partition function (\ref{d9}) in the 
case of supersymmetric as well as fermionic HS spin chain.

It is well known that the spectrum of bosonic $SU(m)$ HS spin chain 
(\ref {a1}) containing $N$ number of lattice sites can be obtained from motifs 
like $\delta \equiv (0, \delta_1, \dots ,\delta_{N-1}, 0)$, 
where each $\delta_j$ is either $0$ or $1$ [6-8]. The form of these 
motifs and corresponding eigenvalues can be reproduced by using
the partition function of bosonic $SU(m)$ HS spin chain [9]. 
Now we want to explore how the motifs associated with $SU(m|n)$
supersymmetric HS spin chain emerge naturally from the expression 
of partition function (\ref{d9}). To this end, 
we define a motif corresponding to the partition ${\bf k}$ 
by using the following rule: $\delta_j=0$ if
$j$ coincides with one of the partial sums $K_i$ and $\delta_j=1$ otherwise. 
Furthermore, it is assumed that 
the lowest power of $q$ in eqn.(\ref{d9}) for the partition 
${\bf k}$ gives the energy eigenvalue $E(\delta)$ of the above motif $\delta$.
In this way we obtain the energy levels of $SU(m|n)$ 
supersymmetric HS spin chain as 
\beq
E(\delta)=
\sum_{i=1}^{r-1}N_i =
\frac{N(N^2-1)}{6} + \sum_{j=1}^{N-1} \delta_j ~j(j-N)\, ,
\label{d10}
\eeq
which apparently coincides with that of the bosonic 
HS spin chain. However it should be observed that, 
for the case of $SU(m)$ spin chain, only those 
${\bf k}$ would contribute in the 
partition function for which $K_{j}-K_{j-1} = k_j \leq m$ [9].
This leads to a selection rule which prohibits the occurrence of 
$m$ or more consecutive 1's within the corresponding motifs. 
On the other hand, since all $\bf{k} \in \mathcal{P}_N$
contribute to the partition function (\ref{d9})
of supersymmetric HS spin chain, 
it is possible to place any number of 
consecutive 1's or 0's within a motif $\delta$. Consequently, 
the selection rule occurring in the bosonic case is lifted for the case 
of supersymmetric HS spin chain and 
many extra energy levels appear in the corresponding spectrum. 
This absence of selection rule 
in the spectrum of supersymmetric HS spin chain was previously 
observed by Haldane on the basis of numerical calculations [3]. 
By using the expression of $E(\delta)$ in eqn.(\ref{d10}), we can easily 
evaluate the maximum and minimum energy eigenvalues of this system. 
From the expression of 
$E(\delta)$ it is evident that,  the motif 
$\delta \equiv (0,0,\dots,0,0)$ would correspond to the maximum 
energy $E_{max}= \frac{N(N^2-1)}{6}$.
 Similarly for the motif $\delta \equiv (0,1,\dots,1,0)$, we 
obtain the minimum energy of the system as
$E_{min}= \frac{N(N^2-1)}{6} + \sum_{j=1}^{N-1} j(j-N)= 0$.
It is interesting to note that these maximum and minimum 
energy eigenvalues of $SU(m|n)$ supersymmetric HS spin chain 
do not depend on the values of $m$ and $n$.  
Moreover, the lifting of the selection rule 
is responsible for the zero minimum energy of supersymmetric 
HS spin chain.

Using Mathematica we find that,  for a wide range of 
values of $m$, $n$ and $N$,  the partition function (\ref{d9}) of $SU(m|n)$ 
HS model satisfies a duality relation of the form
\beq
Z^{(m|n)}(q)=q^{\frac{N(N^2-1)}{6}}Z^{(n|m)}(q^{-1}).
\label{d11}
\eeq
This result motivates us to conjecture that the above duality relation, 
involving the interchange of bosonic and fermionic spin degrees of freedom,
is valid for all possible values of $m$, $n$ and $N$. It may be noted that, 
 for the particular case $n=0$, eqn.(\ref{d11})
 relates the partition function of  $SU(m)$
 bosonic HS spin chain to that of $SU(m)$ fermionic spin chain. 
By applying the relation 
${\tilde P}_{jk}^{(m|0)}= -{\tilde P}_{jk}^{(0|m)}$ and 
the summation formula 
$\sum_{1\leq j <k \leq N} \frac{1}{\sin^2( \xi_j- \xi_k)}=
\frac{N(N^2-1)}{6} $ [9,20], we find that 
 the Hamiltonians of bosonic and fermionic spin chains are connected as 
$$
H^{(m|0)}= \frac{N(N^2-1)}{6} -  H^{(0|m)}  \, .
$$ 
Using the above relation along with the definition of partition function 
given by $Z^{(m|n)}(q)=tr[q^{H^{(m|n)}}]$,
one can easily prove eqn.(\ref{d11}) for the particular case $n=0$. It would 
be interesting  to explore whether eqn.(\ref{d11}) can be also proved
for the general case  by establishing some relation between 
${\tilde P}_{jk}^{(m|n)}$ and ${\tilde P}_{jk}^{(n|m)}$.
Comparing the coefficients of 
same power of $q$ from both sides of eqn.(\ref{d11}), we find that 
the energy levels of $SU(n|m)$ spin chain can be obtained from those of 
$SU(m|n)$  spin chain through the transformation 
$E_i \rightarrow \frac{N(N^2-1)}{6}-E_i $ and also get the relation 
\beq
\mathcal{D}^{(m|n)}\left(E_i\right)
=\mathcal{D}^{(n|m)} \left( \frac{N(N^2-1)}{6}-E_i \right),
\label{d12}
\eeq
where $\mathcal{D}^{(m|n)}(E_i)$ denotes 
the degeneracy factor corresponding to energy $E_i$ of $SU(m|n)$
HS spin chain.  Thus it is evident that, 
the spectrum of $SU(n|m)$ spin chain can be
obtained from that of $SU(m|n)$ spin chain through 
an inversion and overall shift of all energy levels.
Such relation between the spectra of 
supersymmetric HS spin chains was empirically found 
by Haldane with the help of numerical analysis [3].

\noindent \section {Spectral properties of $SU(m|n)$ HS spin chain }
\renewcommand{\theequation}{5.{\arabic{equation}}}
\setcounter{equation}{0}
In this section we shall explore some spectral properties 
of supersymmetric HS model by 
using its exact partition function $Z^{(m|n)}(T)$ (\ref{d9}). 
It has been already mentioned that, 
calculation for the degeneracy factors associated with the energy 
eigenvalues of this spin chain becomes very cumbersome by using the 
motif representations for large values of $N$. However, with the help of
a symbolic software package like Mathematica, it is possible to 
express the partition function (\ref{d9}) as a polynomial of $q$ and
explicitly find out the degeneracy factors of all energy levels for relatively
large values of $N$. In this way,  we can study 
properties like level density distribution and  
nearest-neighbour spacing (NNS) distribution for the spectrum 
of supersymmetric HS spin chain.

For the case of $SU(m)$ bosonic spin chain, it has been found earlier
that the continuous part of the energy level density obeys
Gaussian distribution to a very high degree of accuracy for $N>>1$ [9]. 
At present, our aim is to study the level density distribution in the spectrum
of $SU(m|n)$ supersymmetric HS spin chain and investigate whether it 
exhibits a similar behaviour. To begin with, let us consider
the simplest case of $SU(1|1)$ supersymmetric HS spin chain.  
In this case,  the degeneracy factor in eqn.(\ref{c17})
reduces to a simple form given by 
$ d^{(1|1)}({\bf k})= 2^r$. 
By substituting this degeneracy factor to eqn.(\ref{d9}), taking some  
specific value for the number of lattice sites like $N=15$ and
using Mathematica, we express the partition 
function of $SU(1|1)$ spin chain as a polynomial of $q$. The 
coefficient of $q^{E_i}$ in such polynomial evidently gives the
degeneracy factor  $\mathcal{D}^{(1|1)}(E_i)$ corresponding to the 
energy eigenvalue $E_i$, which we plot in Fig.1. This figure
clearly indicates that the energy level distribution 
obey Gaussian approximation but with some local fluctuations. 
Similar behaviour of energy level distribution has been found by
studying $SU(m|n)$ HS spin chain with other values of $m,~n$ and sufficiently
large values of $N$. 

From the above discussion it is apparent that, if we decompose 
the energy level density associated with $SU(m|n)$ HS spin chain 
as a sum of continuous part and fluctuating part, the continuous part will 
obey Gaussian distribution for large values of $N$. 
This behaviour of the continuous part
can be measured in a quantitative way
 by studying the cumulative level density [9], 
 which eliminates the fluctuating part of the level density distribution. 
For the case of $SU(m|n)$ HS spin chain, 
cumulative level density of the spectrum is defined as
\beq
 F(E)=\frac{1}{(m+n)^N} \sum_{E_i\leq E} \mathcal{D}^{(m|n)} (E_i) \, .
\label {e1}
\eeq
Obviously, this $F(E)$ can also be obtained by expressing 
the exact partition function (\ref{d9}) as a polynomial of $q$. 
We want to check whether this $F(E)$ agrees well  
with the error function given by
\beq
G(E)=\frac{1}{2}\left[1+erf\left(\frac{E-\mu}{\surd{2} \sigma 
}\right)\right],
\label{e2}
\eeq
where $\mu $ and $\sigma $
are respectively the mean value and the standard deviation associated with the 
energy level density distribution. 
These parameters are related to the Hamiltonian $H^{(m|n)}$
(\ref{b9}) as 
\bea
~~~~~~~~~~~~~~~~~~~~\mu=\frac{tr\left[ 
H^{(m|n)}\right]}{{(m+n)}^N}\,, ~~~~~\sigma^2=  \frac{tr\left[(H^{(m|n)})^2 
\right]}{{(m+n)}^N} \, - \, \mu^2. 
~~~~~~~~~~~~~~~~~~~~~(5.3a,b) \nn
\eea

For the purpose of comparing $F(E)$ with $G(E)$,  it is 
necessary to express the parameters $\mu$ and $\sigma$ as some 
functions of $m$, $n$ and $N$. 
To this end, we need the following trace formulas:
\bea
\begin{aligned}
~~~& tr\left[ 
(\tilde{P}_{ij}^{(m|n)})^2\right]=tr\left[ \one \right]=
s^N,~~~~tr\left[ \tilde{P}_{ij}^{(m|n)} 
\right]=s^{N-2}t \, ,
~~~~~~~~~~~~~~~~~~~~~~~~~~~~~~~(5.4 a,b ) \nn \\   
~~~&tr \left[ \tilde{P}_{ij}^{(m|n)}\tilde{P}_{il}^{(m|n)} \right] 
=tr\left[ \tilde{P}_{ij}^{(m|n)}
\tilde{P}_{jl}^{(m|n)}\right] =tr\left[ 
\tilde{P}_{ij}^{(m|n)}\tilde{P}_{kj}^{(m|n)}\right]=s^{N-2},
 ~~~~~~~~~~~~~~~~(5.4 c )  \nn \\
~~~&tr\left[\tilde{P}_{ij}^{(m|n)}\tilde{P}_{kl}^{(m|n)}\right]=s^{N-4}t^2 ,
~~~~~~~~~~~~~~~~~~~~~~~~~~~~~~~~~~~~\,~~~~~~~~~~~~~~~~~~~~~~~~~~~~~(5.4d) \nn
\end{aligned}
\eea
where $s=m+n$, $t=m-n$ and $i,j,k,l$ are all different indices. 
 Derivation of these trace formulas is given 
in Appendix A of this article. For obtaining the functional form of $\mu$ 
and $\sigma$, it is also required to evaluate summations like 
\bea
\begin{aligned}
 ~~~&R_0 \equiv \sum_{i<j} \frac{1}{\sin^2(\xi_i-\xi_j)} \, ,
~~~R_1 \equiv \sum_{i<j} 
\frac{1}{\sin^4(\xi_i-\xi_j)}  ,~~~~~~~~~~~~~~~
~~~~~~~~~~~~~~~~~~(5.5a,b) \nn \\
 ~~~~~&R_2 \equiv \sum_{i<j} 
\sum_{\substack{k<l \\ (k,l \neq i,j)}}
\frac{1}{\sin^2(\xi_i-\xi_j) \sin^2 
(\xi_k-\xi_l)} \, ,
~~~~~~~~~~~~~~~~~~~~~~~~~
~~~~~~~~~~~~~~~~~~~(5.5c) \nn \\
~~~~& R_3 \equiv  2 \sum_{i<j} \sum_{j<l}  
\frac{1}{\sin^2(\xi_i-\xi_j)\sin^2(\xi_j-\xi_l)} 
 \, + \, \sum_{i<j} \sum_{i<l}  
\frac{1}{\sin^2(\xi_i-\xi_j)\sin^2(\xi_i-\xi_l)}  \nn \\
&~~~~+ \, \sum_{i<j}  \sum_{k<j}  
\frac{1}{\sin^2(\xi_i-\xi_j)\sin^2(\xi_k-\xi_j)} \,   .
~~~~~~~~~~~~~~~~~~~~~~~~~~~~~~~~~~~~~~~~~~~~~~~(5.5d) \nn 
\end{aligned}
\eea
It is easy to see that
the above defined $R_0$, $R_1$, $R_2$ and $R_3$ satisfy the relation
\addtocounter{equation}{3}
\beq
R_0^2=R_1+R_2+R_3.
\label{e6}
\eeq
Using some summation formulas given in Ref.20, it can be shown that 
\bea
\begin{aligned}
~~~~~~~~~~~~&R_0=\frac{N(N^2-1)}{6}\, ,
~~~~~~~~~~~~~~~~~~~~~~~~~~~~~~~~~~~~~~~~~~~~
~~~~~~~~~~~~~~~~~~~~~~~~(5.7a) \nn \\
~~~~~~~~~~~~&R_1=\frac{N(N^2-1)(N^2+11)}{90}\, ,
~~~~~~~~~~~~~~~~~~~~\,~~~~~~~~~~
~~~~~~~~~~~~~~~~~~~~~~~~~(5.7b) \nn \\
~~~~~~~~~~~~&R_2=\frac{N(N^2-1)^2(N-4)}{36}+
\frac{N(N^2-1)(N^2+11)}{90} \, ,~~~~~~\,~~~~~~~~~~~~
~~~~~~~(5.7c) \nn \\
~~~~~~~~~~~~&R_3=\frac{4N(N^2-1)(N^2-4)}{45}\,.
~~~~~~~~~~~~~~~~~~~~~~~~~~~~~~~~~~~~~~~~~~~~~
~~~~~~~~~~~(5.7d) \nn 
\end{aligned}
\eea
\addtocounter{equation}{1}
Derivation of these relations is discussed in Appendix B of this article.

Now, by using eqns.(5.3a), (5.4a,b) and (5.7a), we can express
$\mu$ as a function of $m$, $n$ and $N$ given by
\beq
\mu \, = \,  \frac{s^2+t}{2s^2}\, R_0 \, = \, \frac{s^2+t}{12s^2}\,  N(N^2-1).
\label{e8}
\eeq
Next, by using the trace formulas (5.4a,b,c,d), we obtain
\beq
tr\left[(H^{(m|n)})^2\right]\, =\, 
\frac{s^{N-2}\left(s^2+2t\right)}{4} \, R_0^2  + \frac{s^N}{4} \, R_1 
+ \frac{s^{N-4}t^2}{4} \, R_2 + \frac{s^{N-2}}{4} \, R_3 \, .
\label{e9}
\eeq
Substituting the expressions for $\mu$ in eqn.(\ref{e8}) 
and $tr\left[(H^{(m|n)})^2\right]$  in  eqn.(\ref{e9}) 
to eqn.(5.3b), and subsequently using (\ref{e6}), it can be shown that
\beq
\sigma^2 \, = \, \frac{s^4-t^2}{4s^4} \, R_1  + 
\frac{s^2-t^2}{4s^4} \, R_3.
\label{e10}
\eeq
Finally, by substituting the values of $R_1$ 
(5.7b) and $R_3$ (5.7d) to eqn.(\ref{e10}), we can express 
$\sigma$ as a function of $m$, $n$ and $N$ given by
\beq
\sigma \, = \, \left \{ \frac{s^4-t^2}{360s^4} \, N(N^2-1)(N^2+11)
  +   \frac{s^2-t^2}{45s^4} \,  N(N^2-1)(N^2-4) \right \}^\frac{1}{2} \, .
 \label{e11}
\eeq
Thus we are able to find out the functional forms of the 
parameters  $\mu$ and $\sigma$ for the case of $SU(m|n)$ HS spin chain. 
It may be observed that, in the special 
case $n=0$ (for which one gets $s=t=m$), 
eqns.(\ref{e8}) and (\ref{e11}) reproduce the forms of
 $\mu$ and $\sigma$ corresponding to the $SU(m)$ bosonic
HS spin chain [9]. 

Now we can compare the cumulative level density $F(E)$ (\ref {e1}) 
with the error function $G(E)$ (\ref{e2}), where the values of 
$\mu$ and $\sigma$ are obtained from 
eqns.(\ref{e8}) and  (\ref{e11}) respectively 
for any given $m$, $n$ and $N$.  
In Fig.2, we plot such $F(E)$
and $G(E)$  for the particular 
case of $SU(1|1)$ spin
 chain with $N=15$ lattice sites. From this figure 
it is evident that $F(E)$ follows
$G(E)$ to a high degree of approximation. One can
also quantify the agreement between $F(E)$ and
 $G(E)$ by calculating the 
corresponding mean square error (MSE), which for the above mentioned 
case is given by $8.46 \times 10^{-6}$. 
It may be noted that, the agreement between $F(E)$ and $G(E)$ improves 
rapidly with increasing values of $N$. For example, 
in the case of $SU(1|1)$ model,
MSE between $F(E)$ and $G(E)$ decreases from $5.17 \times 10^{-5}$ 
to $8.46 \times 10^{-6}$ when the value of $N$ is increased from $10$ to $15$.
Next, we consider the particular cases of $SU(1|2)$ as well as 
$SU(2|1)$ supersymmetric spin chain with $N=15$ lattice sites and 
also the $SU(3)$ bosonic spin chain with same number of lattice sites 
for the sake of comparison. 
In Fig.3, we plot 
$F(E)$ and $G(E)$ for $SU(1|2)$, $SU(2|1)$ and 
$SU(3)$ HS spin chains and find that the
corresponding MSEs are given by  $2.38 \times 10^{-5}$,
$1.2 \times 10^{-5}$ and $8.61 \times 10^{-6}$ respectively.
Again $F(E)$ shows very good agreement with $G(E)$ for all of these cases. 
Such agreement also improves rapidly with increasing values of $N$.  
For example, in the case of $SU(1|2)$ spin chain,
MSE between $F(E)$ and $G(E)$ decreases from
$8.23 \times 10^{-5}$  to $2.38 \times 10^{-5}$ 
when the value of $N$ is increased from $10$ to $15$. Analysing 
many other particular cases with different values of $m, ~n$
and sufficiently large values of $N$, we find that 
$F(E)$ follows $G(E)$ with a high degree of approximation 
for all of these cases. 

From the above discussion it is evident that 
the local fluctuations in energy level distribution,
 as shown in Fig.1 for the particular
case of $SU(1|1)$ spin chain, get cancelled very
 rapidly whenever we take the cumulative sum of such 
 distribution.  Furthermore, for sufficiently large values of $N$, 
  continuous part of the level density distribution 
 in the spectrum of supersymmetric HS spin chain
satisfies the Gaussian approximation at the same high level of accuracy 
as in the pure bosonic case. 
It may be noted that, the level density of embedded Gaussian orthogonal 
ensemble (GOE) also follows Gaussian distribution at the limit 
$N \rightarrow \infty$, provided the number of one-particle states tends 
to infinity faster than $N$ [21].  However, in our case of $SU(m|n)$ HS spin 
chain, the number of one-particle states (i.e., $m+n$ ) remains fixed for 
all values of $N$.

Next we want to study the NNS distribution in the 
spectrum of $SU(m|n)$ HS spin chain. 
To eliminate the effect of level density variation in the
calculation of NNS distribution for the full energy range, 
it is necessary to apply an unfolding mapping to the `raw' spectrum [22].
This unfolding mapping may be defined by using the continuous part of the
cumulative level density distribution. We have already seen that,
for the case of $SU(m|n)$ HS spin chain, the continuous part of cumulative
level density is given by $G(E)$ (\ref{e2}) with a high degree of approximation. 
So we transform each energy $E_i$, $i=1,\cdots ,l \, $, into an
 unfolded energy $\xi_i \equiv  G(E_i)$. The function $p(u)$ 
is defined as the density of the normalized 
spacings $u_i=(\xi_{i+1}-\xi_i)/\Delta$, where 
$\Delta=(\xi_l-\xi_1)/(l-1)$ is the mean spacing 
of the unfolded energy. To get rid of local fluctuations occurring in $p(u)$,
again we study the cumulative NNS
distribution given by $P(u)= \int_0^u p(x)dx$, instead of $p(u)$.
 In this context it may be noted that,
 NNS distributions corresponding to the cases of
classical GOE as well as embedded GOE obey the Wigner's law [23]:
\beq
p(u)= \frac{\pi}{2} u \exp (-\pi u^2/4) \, . \nn
\eeq
On the other hand, from the conjecture of Berry and Tabor one may 
expect that the NNS distribution for an integrable model will
obey Poisson's law given by $p(u)= \exp(-u)$ [24]. However, it has been
found that the NNS distribution for $SU(m)$ bosonic 
HS spin chain does not follow either
 Wigner's law or Poisson's law within a wide range
 of $N$ [9]. Instead, the cumulative NNS distribution
for this bosonic spin chain can be well approximated 
by a function like
\beq
\tilde{P}(u)=v^{\alpha} [ 1- \gamma (1-v)^{\beta} ],
\label{e12}
\eeq
where $v=u/u_{max}$ with $u_{max}$ being the largest normalized spacing, 
 $\alpha$ and $\beta$ are
two free parameters taking values within the range
$0<\alpha,~\beta<1$, and the value of $\gamma$ is fixed by requiring that 
the average normalized spacing be equal to 1.
In our study we also find that,  NNS distribution 
in the spectrum of $SU(m|n)$ HS spin chain does not follow either
 Wigner's law or Poisson's law within a wide range of $N$. 
In particular, it is observed that the slope of cumulative NNS distribution
diverges for both $u \rightarrow 0$ and
 $u \rightarrow u_{max}$, which can not be explained 
from Wigner's or Poisson's distribution. Furthermore, we find 
that the cumulative NNS distribution for $SU(m|n)$ HS spin chain 
can be fitted well by $\tilde{P}(u)$ in eqn.(\ref{e12}) within a 
range of $N$.
For example, in the particular case of $SU(1|1)$ spin chain,  it is
checked that $P(u)$ agrees well with $\tilde{P}(u)$ 
within the range $N\leq 20$. In Fig.4, 
we plot such $P(u)$ and $\tilde{P}(u)$ for 
$N=17$ lattice sites ($u_{max}=2.626$ in this case) and found a 
good agreement with  MSE $=0.0234$ when 
the values of free parameters are taken as  $\alpha=0.39$ and $\beta=0.29$.
However, it is possible that the appearance of such non-Poissonian 
NNS distribution in the spectrum of $SU(m|n)$ HS spin chain is an
artifact of finite-size effect, which requires further
investigation.

Finally we want to make a comment about the behaviour of 
parameters $\mu$ in eqn.(\ref{e8}) and $\sigma$ in eqn.(\ref{e11})
under the exchange of bosonic and fermionic spin degrees of freedom.
Since  $s \rightarrow s$ and $t \rightarrow -t$ under this exchange,  
we find that $\sigma$ remains invariant and
$\mu$ changes to $\bar{\mu}$ given by
 $\bar {\mu}=\frac{N(N^2-1)}{6} - {\mu} \,$. 
It is interesting to note that this relation between $\mu $ and 
$\bar {\mu}$ can also be obtained by applying eqn.(\ref{d12}): 
\bea
{\mu} &=&\frac{1}{s^N} \sum_{E_i}\mathcal{D}^{(m|n)} (E_i) E_i  \nn \\
&=&\frac{1}{s^N} \sum_{E_i}\mathcal{D}^{(n|m)} 
\left(\frac{N(N^2-1)}{6}-E_i \right) 
E_i 
=\frac{N(N^2-1)}{6}-\bar{\mu} \, . \nn
\eea
This agreement clearly gives a support to our conjecture
(\ref{d11}).  By using this conjecture we have found in Sec.4 that, 
 the spectrum of $SU(n|m)$ spin chain can be
 obtained from that of $SU(m|n)$ spin chain through 
an inversion and overall shift of all energy levels.
Since none of these operations change the
standard deviation of level density distribution, $\sigma$
should take the same value for $SU(m|n)$ and $SU(n|m)$ 
HS spin chain. Hence, the observation that 
$\sigma$ in eqn.(\ref{e11}) remains invariant 
under the exchange of bosonic and fermionic spin degrees of freedom, 
is also consistent with our conjecture (\ref{d11}).

\noindent \section {Conclusion }

Here we derive an exact expression for the partition function
of $SU(m|n)$ supersymmetric HS spin chain by using the freezing trick
and also study some properties of the related spectrum.
For applying the freezing trick, we consider a spin CS model containing 
an anyon like representation of the permutation algebra as spin dependent 
interaction. We find out the complete spectrum of such spin CS model
including the degeneracy factors of all energy eigenvalues.
At the strong coupling limit, this spin CS model 
reduces to the sum of spinless CS model with only dynamical degrees
of freedom and $SU(m|n)$ supersymmetric HS spin chain. 
Consequently, by factoring out the contribution due to dynamical
degrees of freedom from partition function of this spin CS model, 
we obtain the partition function of $SU(m|n)$ supersymmetric HS spin chain. 
By using this partition function, we study the motif representation for 
$SU(m|n)$ HS spin chain and find that, due to the lifting 
of a selection rule, some additional energy levels appear in the 
spectrum in comparison with the case of $SU(m)$ bosonic spin chain. 

By using Mathematica we observe that,  the partition function of 
$SU(m|n)$ HS model satisfies the duality relation (\ref{d11})
for many values of $m,$ $n$ and $N$. This observation 
motivates us to conjecture that this duality relation, involving 
the interchange of bosonic and fermionic spin degrees of freedom,
is valid for all possible values of $m,$ $n$ and $N$.  It would 
be interesting if this duality relation can be proved 
analytically by using the motif representations and skew-Young 
diagrammes associated with the $Y(gl_{(m|n)})$ quantum group.  
Furthermore, it is known that, 
the partition functions of $SU(m)$ and $SU(m|n)$ 
Polychronakos spin chains are intimately connected with 
Rogers-Szeg\"{o} (RS) polynomial [8,15],
which appears in the theory of partitions [25].
Since, HS spin chain share the same quantum 
group symmetry with Polychronakos spin chain, 
it might be promising to investigate mathematical
structures connected with the partition functions of $SU(m)$ as well as
$SU(m|n)$ HS spin chain and explore whether some 
new RS type polynomials can be generated in this way. 

By using the partition function of $SU(m|n)$ HS spin chain, 
we study global properties of its spectrum like level
density distribution and  NNS distribution.
It is found that, similar to the case of $SU(m)$ bosonic HS spin chain,
 continuous part of the energy level density
satisfies the Gaussian distribution with a high degree of accuracy
for sufficiently large values of $N$.   
We also derive exact expressions for the mean value and the
standard deviation which characterize such 
Gaussian distribution.  It would be interesting to provide 
an explanation for this behaviour of energy level density distribution 
in the framework of random matrix theory
and explore whether the underlying quantum group symmetry 
of HS spin chain plays some role in this matter. 

\vskip 1 cm
\noindent{\bf Acknowledgements}
\smallskip

We would like to thank Palash B. Pal for some helpful discussions. 

\newpage

\newpage
\renewcommand{\theequation}{A-\arabic{equation}}
\setcounter{equation}{0}
\section*{\begin{large} Appendix A. 
\end{large}\begin{normalsize} Evaluation of trace formulas \end{normalsize}} 

Here we shall derive the trace formulas (5.4a,b,c,d)
 by assuming that $\v \alpha_1 \dots \alpha_N \r$
(with $\alpha_j \in [1,\dots,s]$) are orthonormal 
set of vectors.  Since the trace of identity operator is given by the dimension 
of the Hilbert space, eqn.(5.4a) is really a trivial relation.
Using eqn.(\ref{b6}), it can be shown that
\bea
\l\alpha_1 \dots \alpha_i \dots \alpha_j \dots \alpha_N 
\v  \tilde{P}_{ij}^{(m|n)}
\v \alpha_1 \dots \alpha_i \dots \alpha_j \dots \alpha_N \r
=(-1)^{\epsilon (\alpha_i) } \delta_{\alpha_i\alpha_j} , \nn
\eea
where $\epsilon(\alpha_i)=0~(1)$ when $\alpha_i$ is a
 bosonic (fermionic) spin.
With the help of above equation, we derive 
the trace relation (5.4b) as 
\bea
tr \left[\tilde{P}_{ij}^{(m|n)}\right]
\hskip -.20 cm &=& \hskip -.20 cm 
\sum_{\alpha_1\dots\alpha_N=1}^{s }
 \l \alpha_1 \dots  \alpha_i \dots \alpha_j \dots 
\alpha_N \v 
\tilde{P}_{ij}^{(m|n)}\v \alpha_1 \dots  \alpha_i 
\dots \alpha_j \dots \alpha_N  \r  \nn \\
&=& ~ \sideset{}{^{(\alpha_i,\alpha_j)}}\sum_{\alpha_1
\dots\alpha_N=1}^{s }
\sum_{\alpha_i,\alpha_j=1}^s  \, \l \alpha_1 \dots  
\alpha_i \dots \alpha_j \dots \alpha_N \v 
\tilde{P}_{ij}^{(m|n)}\v \alpha_1 \dots  \alpha_i 
\dots \alpha_j \dots \alpha_N  \r  \nn \\
&=& ~ \sideset{}{^{(\alpha_i,\alpha_j)}}\sum_{\alpha_1
\dots\alpha_N=1}^{s}
\sum_{\alpha_i=1}^s (-1)^{\epsilon 
(\alpha_i)}=s^{N-2}t , \nn 
\eea
where the notation
$\sideset{}{^{(\alpha_i,\alpha_j)}}\sum\limits_{\alpha_1 \dots\alpha_N=1}^{s}$
represents summation  over all spin components 
$\alpha_1,\dots,\alpha_N$ except $\alpha_i$ and
$\alpha_j$. 

Next, by using eqn.(\ref{b6}), it is found that
\bea
~~~~~~~~~~\l\alpha_1 \dots  \alpha_i \dots \alpha_j 
\dots \alpha_l \dots\alpha_N \v \tilde{P}_{ij}^{(m|n)} 
\tilde{P}_{il}^{(m|n)} 
\v \alpha_1 \dots  \alpha_i \dots \alpha_j 
\dots \alpha_l \dots \alpha_N \r 
=  \delta_{\alpha_i\alpha_j} \delta_{\alpha_i\alpha_l}.\nn
\eea
Applying the above equation, we obtain 
 a trace relation in eqn.(5.4c) as 
\begin{equation}
\begin{split}
& tr\left[ \tilde{P}_{ij}^{(m|n)}\tilde{P}_{il}^{(m|n)} 
\right] \\
&~~~~ =\sum_{\alpha_1\dots\alpha_N=1}^{s } 
\l \alpha_1 \dots \alpha_i \dots \alpha_j 
\dots \alpha_l \dots \alpha_N \v 
\tilde{P}_{ij}^{(m|n)}\tilde{P}_{il}^{(m|n)} 
\v \alpha_1 \dots \alpha_i \dots \alpha_j 
\dots \alpha_l \dots  \alpha_N \r \nn \\
&~~~~=~~\sideset{}{^{(\alpha_i,\alpha_j,\alpha_l)}}
\sum_{\alpha_1\dots\alpha_N=1}^{s } 
\sum_{\alpha_i=1}^s 1
=s^{N-2}.
\end{split}
\end{equation}
 Other trace relations in (5.4c) can be proved in a similar way.

By using eqn.(2.6), it is also found that
\bea
~~~~~~&\l \alpha_1 \dots \alpha_i \dots \alpha_j \dots 
\alpha_k \dots \alpha_l \dots \alpha_N \v \, 
\tilde{P}_{ij}^{(m|n)}\tilde{P}_{kl}^{(m|n)} \, \v
 \alpha_1 \dots \alpha_i \dots \alpha_j \dots 
\alpha_k \dots \alpha_l \dots \alpha_N \r ~~~~~~~~~~~~~~~\nn \\
&~~~~~~~~~~~~~~~~~~~~~~~~=(-1)^{\epsilon(\alpha_i)+\epsilon(\alpha_k)} 
\, \delta_{\alpha_i \alpha_j} \delta_{\alpha_k 
\alpha_l}\, . \nn
\eea
With the help of this equation, we obtain the trace relation (5.4d) as
\begin{equation}
\begin{split}
&tr\left[ \tilde{P}_{ij}^{(m|n)}\tilde{P}_{kl}^{(m|n)} 
\right] \\
&=\sum_{\alpha_1\dots\alpha_N=1}^{s }
\l\alpha_1 \dots \alpha_i \dots \alpha_j 
\dots \alpha_k \dots \alpha_l \dots \alpha_N\v
\tilde{P}_{ij}^{(m|n)}\tilde{P}_{kl}^{(m|n)}
\v \alpha_1 \dots \alpha_i \dots \alpha_j \dots 
\alpha_k \dots \alpha_l \dots \alpha_N \r   \\
&=~~\sideset{}{^{(\alpha_i,\alpha_j,\alpha_k,
\alpha_l)}}\sum_{\alpha_1\dots\alpha_N=1}^s 
\sum_{\alpha_i ,\alpha_k=1}^s
(-1)^{\epsilon (\alpha_i)+ \epsilon(\alpha_k)}= s^{N-4}t^2.  \nn
\end{split}
\end{equation}

\renewcommand{\theequation}{B-\arabic{equation}}
\setcounter{equation}{0}
\section*{\begin{large} Appendix B. \end{large} \begin{normalsize}                   
              Evaluation of summation formulas\end{normalsize}} 
Here we briefly describe the way of calculating known
summation formulas (5.7a) and 
(5.7b) [9],  and subsequently present our derivation 
for new ones like (5.7c) and (5.7d). 
From the work of Calogero et. al [20], it is 
 known that
\beq
\sum_{j=1}^{N-1} \frac{1}{\sin^2 ( \frac{j \pi}{N})}= 
\frac{N^2-1}{3}\, ,
\label{B1}
\eeq
and
\beq
\sum_{j=1}^{N-1} \frac{1}{\sin^4 (\frac{j \pi }{N})}= 
\frac{(N^2-1)(N^2+11)}{45}\, .
\label{B2}
\eeq
Using the translational invariance on a 
circular lattice and summation relation (\ref {B1}),
one can obtain eqn.(5.7a) as  
\beq
R_0=\frac{1}{2}\sum_{i\neq j} \frac{1}{\sin^2(\xi_i-\xi_j)}
=\frac{N}{2}\sum_{j=1}^{N-1} \frac{1}{\sin^2 \left(\frac{j\pi}{N}\right)}
=\frac{N(N^2-1)}{6}\,. \nn
\eeq
Similarly, by using (\ref{B2}), one obtains eqn.(5.7b) as 
\beq
R_1=\frac{1}{2}\sum_{i\neq j} \frac{1}{\sin^4(\xi_i-\xi_j)}
=\frac{N}{2}\sum_{j=1}^{N-1} \frac{1}{\sin^4 \left(\frac{j\pi}{N}\right)}
=\frac{N(N^2-1)(N^2+11)}{90}. \nn
\eeq

For the purpose of calculating  $R_2$ in eqn.(5.5c), 
we note that $R_0$ in eqn.(5.5a) can be expressed as
\beq
R_0= \sum_{\substack{k<l \\ (k,l \neq i,j)}}
\frac{1}{\sin^2(\xi_k-\xi_l)}+
\sum_{\substack{r=1 \\ (r \neq i)}}^N \frac{1}{\sin^2(\xi_i-\xi_r)} 
+ \sum_{\substack{r=1 \\ (r \neq j)}}^N  \frac{1}{\sin^2(\xi_j-\xi_r)}
- \frac{1}{\sin^2(\xi_i-\xi_j)}. \nn
\eeq
Substituting the value of $R_0$ given in (5.7a) to the above relation and
also using (\ref {B1}), we find that 
\beq
\sum_{\substack{k<l \\ (k,l \neq i,j)}} \frac{1}{\sin^2(\xi_k-\xi_l)}
=\frac{(N^2-1)(N-4)}{6} 
+ \frac{1}{\sin^2(\xi_i-\xi_j)}\, .  \nn
\eeq
By substituting this expression 
to $R_2$ in  eqn.(5.5c), and subsequently using eqns.(5.7a) as well as 
(5.7b), we derive the value of $R_2$ given in eqn.(5.7c) as 
\bea
R_2 &=& \sum_{i<j} \frac{1}{\sin^2(\xi_i-\xi_j)} 
\biggl[\frac{(N^2-1)(N-4)}{6} + 
\frac{1}{\sin^2(\xi_i-\xi_j)} \biggr] \nn \\
&=& \frac{N(N^2-1)^2(N-4)}{36} + \frac{N(N^2-1)(N^2+11)}{90} \,. \nn
\eea
Finally, by substituting the values of  $R_0$ (5.7a), 
$R_1$ (5.7b) and $R_2$ (5.7c) to the relation (\ref{e6}), we 
easily obtain the value of $R_3$  appearing in eqn.(5.7d).

\newpage
\vskip 2 cm 
\leftline {\large \bf References}
\medskip
\begin{enumerate}

\item F. D. M. Haldane, Phys. Rev. Lett.  60 (1988) 635.
\item B. S. Shastry, Phys. Rev. Lett.  60 (1988) 639.
\item F. D. M Haldane, in Proc. 16th Taniguchi Symp., Kashikojima, Japan 
(1993), eds.  A. Okiji and N. Kawakami (Springer, 1994).
\item Z.N.C. Ha, {\it Quantum many-body systems in one dimension} 
(Series on Advances in Statistical Mechanics, Vol.12), 
(World Scientific,1996). 
\item A.P. Polychronakos,  {\it Generalized statistics in one dimension}, 
Les Houches 1998 lectures, hep-th/9902157.
\item F. D. M. Haldane, Z.N.C. Ha, J.C. Talstra, 
 D. Benard and V. Pasquier, Phys. Rev. Lett. 69 (1992) 2021. 
\item D. Benard, M. Gaudin, F. D. M. Haldane, and V. Pasquier, 
J. Phys.  A26 (1993) 5219.
\item  K.Hikami, Nucl. Phys. B441[FS] (1995) 530.
\item F. Finkel, A. Gonz\'{a}lez-L\'{o}pez, Phys. Rev. 
 B72 (2005) 174411.
\item A.P. Polychronakos, Phys. Rev. Lett. 70 (1993) 2329; Nucl. Phys. 
B419 (1994) 553.
\item Bill Sutherland and B. S. Shastry, Phys. Rev. Lett. 71 (1993) 5.
\item A.P. Polychronakos, 
{\it Physics and mathematics of Calogero particles,} hep-th/0607033.
\item P. Schlottmann, Int. Jour. Mod. Phys.  B11 (1997) 355. 
\item B. Basu-Mallick, H. Ujino and M. Wadati, Jour. Phys. Soc. Jpn. 68
(1999) 3219. 
\item K. Hikami, and B. Basu-Mallick, Nucl. Phys.  B566
[PM] (2000) 511.
\item B. Basu-Mallick, Nucl. Phys.  B540 [FS] (1999) 679.
\item B. Basu-Mallick, Nucl. Phys.  B482 [FS] (1996) 713.
\item C. Ahn and W. M. Koo, Phys. Lett. B365 (1996) 105. 
\item B. Sutherland, Phys. Rev. A5 (1972) 1372. 
\item F. Calogero and A.M. Perelomov, Commun. Math. Phys. 59 (1978) 109.
\item K.K. Mon and J.B. French, Ann. Phys. 95 (1975) 90. 
\item F. Haake, {\it Quantum signatures of Chaos}
 (Springer-verlag, 2001).
\item V.K.B. Kota, Phys. Rep. 347 (2001) 223. 
\item M.V. Berry and M. Tabor, Proc. R. Soc. Lond. A356 (1977) 375. 
\item G.E. Andrews, {\it The theory of partitions}
      (Addison-Wesley, Reading, M.A., 1976).

\end{enumerate}

\newpage

\begin {figure} [h]
\centering
\includegraphics[scale=0.87]{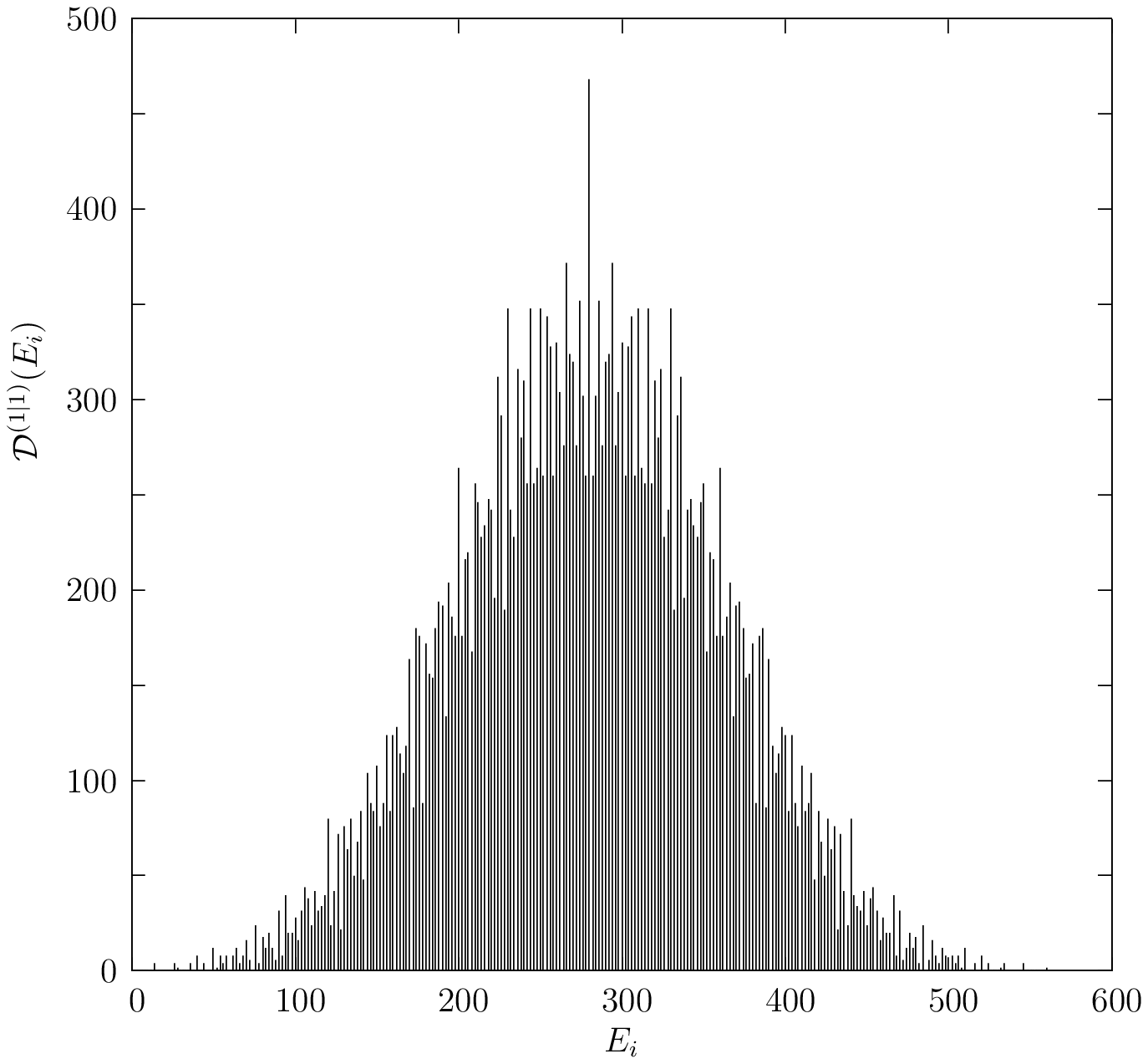}
\caption{Energy levels $E_i$ and degeneracies $\mathcal{D}^{(1|1)}(E_i)$ 
of the $SU(1|1)$ HS spin chain for $N=15$.}
\end{figure}
\newpage
\begin {figure} [h]
\hskip -.25 cm 
\includegraphics[scale=0.65,angle=270]{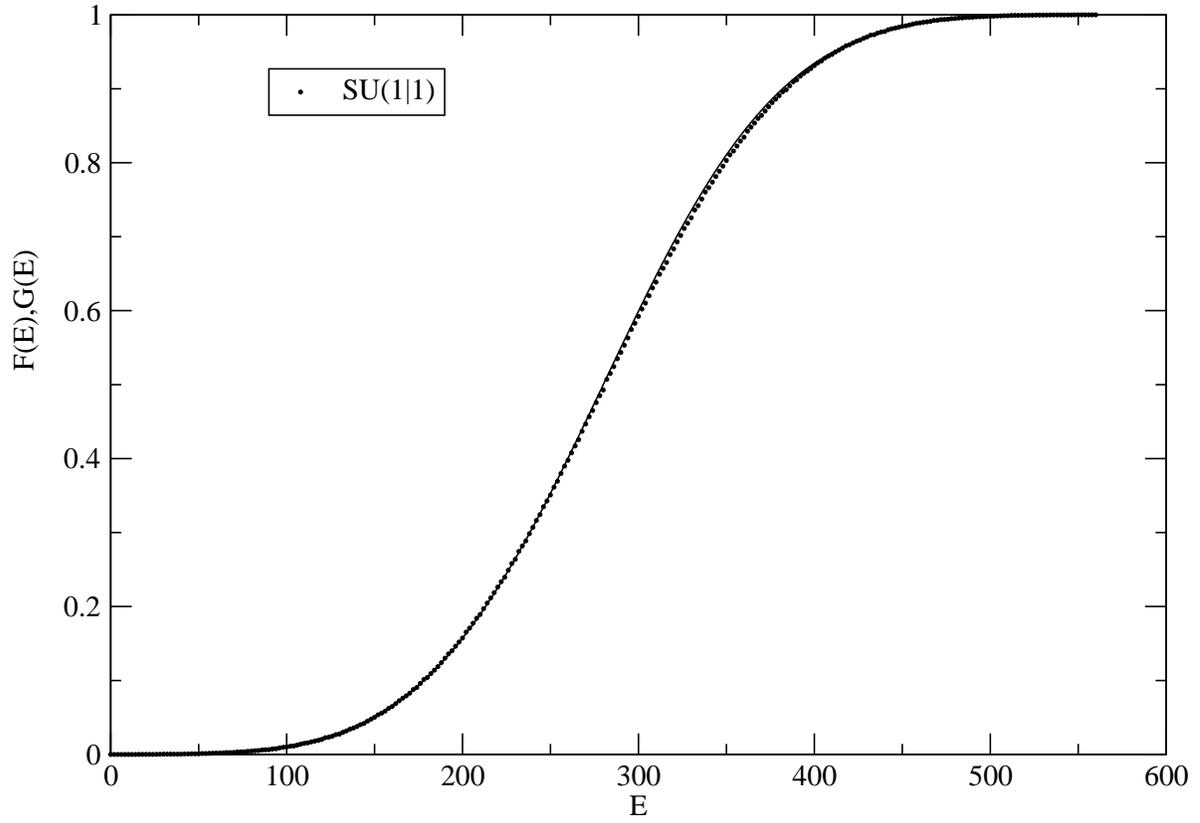}
\caption{Continuous curve represents the error function G(E) 
and crosses represent the 
cumulative distribution function F(E) (at its discontinuity points)
for $SU(1|1)$ spin chain with $N=15$.} 
\end{figure}
\newpage
\begin {figure} [h]
\centering
\includegraphics[scale=0.60,angle=270]{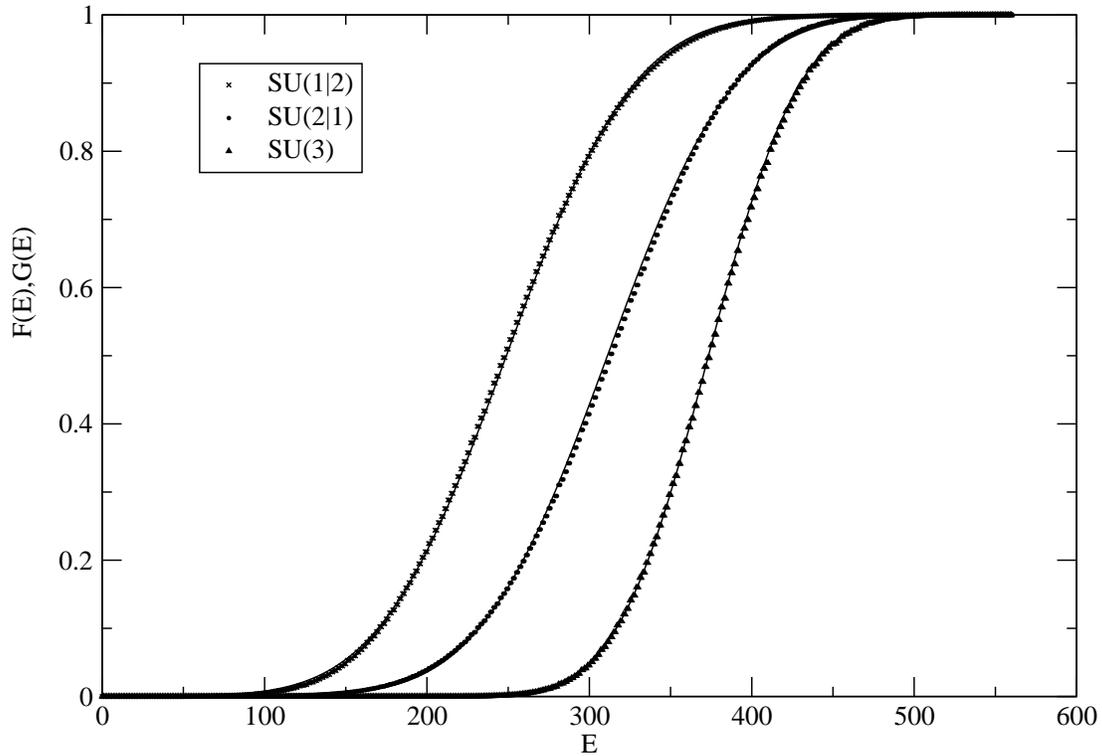}
\caption{Left continuous curve represents G(E) for $SU(1|2)$ 
spin chain with $N=15$ and the corresponding F(E) 
is given by crosses. Middle continuous curve 
represents G(E) for $SU(2|1)$ spin chain
with $N=15$ and the corresponding F(E) is given by dots.
Right continuous curve represents G(E) for $SU(3)$ spin chain
with $N=15$ and the corresponding F(E) is given by triangles.}
\end{figure}
\newpage
\begin {figure} [h]
\centering
\includegraphics[scale=0.60,angle=270]{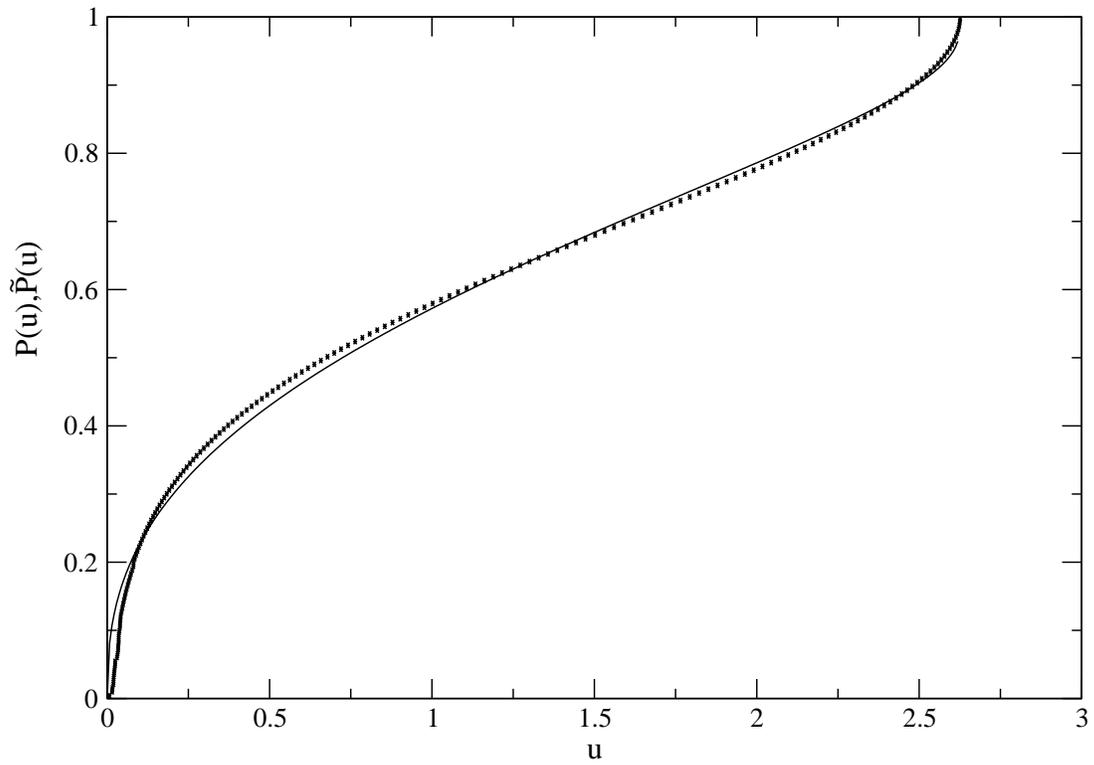}
\caption{Dotted curve represents the cumulative NNS distribution
$P(u)$ for $SU(1|1)$ spin chain with $N=17$ and the continuous
curve represents the approximate distribution function $\tilde{P}(u)$.}
\end{figure}

\end{document}